\documentclass[journal]{IEEEtran}
\usepackage{graphicx}
\usepackage{amsmath}
\usepackage{amssymb}
\usepackage{afterpage}
\usepackage{verbatim}
\usepackage{psfrag}
\usepackage{color}
\usepackage{cite}
\usepackage{setspace}
\usepackage{epsfig}
\usepackage{epstopdf}
\usepackage{footmisc}
\usepackage{fmtcount}
\usepackage{stackrel}
\usepackage{float}
\usepackage{breqn}
\usepackage{enumerate}
\usepackage{graphicx}
\usepackage{subcaption}
\begin{document}

\title{A State-of-the-Art Survey on Multidimensional Scaling Based Localization Techniques}

\author{Nasir Saeed,~\IEEEmembership{Senior Member,~IEEE}, Haewoon Nam,~\IEEEmembership{Senior Member,~IEEE},  Tareq Y. Al-Naffouri,~\IEEEmembership{Senior Member,~IEEE}, Mohamed-Slim Alouini,~\IEEEmembership{Fellow,~IEEE}
\thanks{This research was supported by Basic Science Research Program through the National Research Foundation of Korea (NRF)
funded by the Ministry of Education (2016R1D1A1B03934277). This work was also supported by the Office of Sponsored Research (OSR) at King Abdullah University of Science and Technology (KAUST).
(Corresponding author: Haewoon Nam.)}
\thanks{N.~Saeed, T.~Y.~Al-Naffouri, and M.-S.~Alouini are with the Department of Electrical Engineering, Computer Electrical and Mathematical Sciences \& Engineering (CEMSE) Division, King Abdullah University of Science and Technology (KAUST), Thuwal, Makkah Province, Kingdom of Saudi Arabia, 23955-6900.}
\thanks{H. Nam is with the Division of Electrical Engineering, Hanyang University, Ansan, 15588, Korea.}
}

\maketitle{}
\begin{abstract}
Current and future wireless applications strongly rely on precise real-time localization. A number of applications such as smart cities, Internet of Things (IoT), medical services, automotive industry, underwater exploration, public safety, and military systems require reliable and accurate localization techniques. Generally, the most popular localization/ positioning system is the Global Positioning System (GPS). GPS works well for outdoor environments but fails in indoor and harsh environments. Therefore, a number of other wireless local localization techniques are developed based on terrestrial wireless networks, wireless sensor networks (WSNs) and wireless local area networks (WLANs). Also, there exist localization techniques which fuse two or more technologies to find out the location of the user, also called signal of opportunity based localization. Most of the localization techniques require ranging measurements such as time of arrival (ToA), time difference of arrival (TDoA), direction of arrival (DoA) and received signal strength (RSS). There are also range-free localization techniques which consider the proximity information and do not require the actual ranging measurements. Dimensionality reduction techniques are famous among the range free localization schemes. Multidimensional scaling (MDS) is one of the dimensionality reduction technique which has been used extensively in the recent past for wireless networks localization. In this paper, a comprehensive survey is presented for MDS and MDS based localization techniques in WSNs, Internet of Things (IoT), cognitive radio networks, and 5G networks.
\end{abstract}


\maketitle

\IEEEdisplaynotcompsoctitleabstractindextext

\IEEEpeerreviewmaketitle
\section{Introduction}
Accurate, real-time and reliable localization systems are required for the future generation of wireless communication networks \cite{Junglas2008}. Localization systems enable a user to find its location, and make use of the location for location-based services (LBS) such as monitoring \cite{Camp}, tracking, and navigating \cite{Gartner}, etc. The performance of wireless networks is significantly improved with the addition of location information for network planning \cite{Rao2003}, resource allocation \cite{Nam2015},  load balancing \cite{Yanmaz2005}, spatial spectrum sensing \cite{Nasir2015}, and network adaptation \cite{Bush2005}, etc. Global positioning systems are also known as global navigation and satellite systems (GNSS) allow each user to figure out its location globally. GNSS consists of different positioning systems from different countries such as the global positioning system (GPS), GALILEO, "Globalnaya navigatsionnaya sputnikovaya sistema" (GLONASS) and BeiDou \cite{teng_wang_2016}. GPS and GNSS work well for outdoor environments, but it fails to localize a user in an indoor or harsh environment. In comparison to the outdoor environment, the indoor environment is more challenging and complex. The various obstacles such as human beings, walls, equipment's, ceilings, etc., influence the propagation of signals, thus leads to multi-path propagation error. In addition to that, interference is also added to the propagating signal by noise sources from other wireless networks. Considering these issues in the indoor environment, the development of indoor positioning systems is challenging for future wireless communication systems.

A number of survey articles are presented on the design and development of indoor positioning systems such as \cite{Zhu2014}, \cite{Ramon2017}, and \cite{Faheem2017}. Indoor positioning systems have been developed by different research centers, companies, and universities based on various wireless communication technologies operating on different frequencies such as acoustic waves, radio frequency (RF), ultra-wideband, infrared, and visible light. All of the above mentioned indoor positioning systems are based on a specific ranging technique.

Since the cost and hardware limitation of sensors often prevent the localization systems from using range-based techniques, range-free localization techniques are developed to substitute the range-based techniques. Range-free localization techniques are dependent on the connectivity information which is a much cheaper solution than the range-based techniques because these techniques do not require extra hardware to compute the actual range and rely only on the proximity information \cite{SINGH2015, Ahmadi2016}. Range-free schemes are performed by using the constraint optimization, geometric interpretation, and area formation techniques \cite{Stoleru2007}. Multidimensional scaling (MDS) is one of the most common network localization techniques which can work for both range-free and range-based schemes.

MDS is one of the dimensionality reduction techniques which converts multidimensional data into lower dimensional space while keeping the essential information. The main benefit of using MDS is to get
a graphical display for the given data, such that it is much easier to understand. There exists other dimensionality reduction techniques like principal component analysis (PCA), factor analysis and Isomap but MDS is much popular among all these techniques because of its simplicity and many application areas. MDS analysis finds the spatial map for objects given that the similarity or dissimilarity information between the objects is available \cite{Borg}.

In the recent past, MDS is widely used for localization and mapping of wireless sensor networks (WSNs) and the internet of things (IoT).  In \cite{Shang} a proximity information based sensor network localization is proposed, where the main idea is to construct a local configuration of sensor nodes by using classical MDS (CMDS). The MDS based localization algorithms in \cite{Shang} and \cite{viv} are centralized with higher computational complexity \cite{Nasir2015}.    Semi-centralized (or clustered) MDS techniques are developed to compute local coordinates of nodes, which then are refined to find the final position of the nodes \cite{Gwo, Minhan}.   In \cite{patwari}, \cite{Shanchnag} and \cite{Yin} the authors proposed manifold learning to estimate the sensor nodes position in wireless sensor networks. In \cite{Macagano} the authors proposed Nystrom approximation for the proximity information matrix in MDS to reduce its size for better localization accuracy in sensor networks. Distributed MDS based localization algorithm is proposed in \cite{Minghu} with noisy range measurements, where the authors assume that the distances are corrupted with independent Gaussian random noise. MDS methods with different refinement schemes have also been proposed in the literature to get better localization accuracy for the sensor nodes in WSNs  \cite{Latsu, Zhu, nasir2}. More recently a Euclidean distance matrix completion method is proposed for MDS in \cite{Nguyen2016, Nguyen2017} to find the map of an IoT network. Although the literature on MDS based network localization techniques is not rich, it can be well adapted for modern wireless communication systems such as Internet of things IoT, 5G networks, and underwater wireless communication networks.  MDS based network localization can provide efficient data fusion mechanisms for IoT networks. Similarly, MDS based location awareness for 5G networks will provide numerous applications such as radio resource management, routing, and defining radio maps. Moreover, MDS based localization for software-defined networks will enable a centralized map of the whole network including the different entities of the system which can be helpful for various networking issues. In short, all of the modern wireless communication networks require accurate network localization schemes to provide different applications which include but not limited to data tagging, location-aware routing, environment monitoring, and navigation. Therefore, MDS is one of the famous network localization technique which can be applied to these networks to provide such applications.

\subsection{Related Surveys}
A quite good number of survey articles have been presented on the subject of localization systems where the focus of each survey is either narrow or outdated by the technological advancement \cite{Ijaz2013, Hakan2010, Liu2007, Gu2009, Ammar2014, Vo2016, Yassin2016, Ferreira2017}. For example, the survey in \cite{Ijaz2013} is only focused on ultrasonic localization techniques, whereas the works presented in \cite{Hakan2010, Liu2007, Gu2009} are outdated for current technologies although their goals remain unchanged. In \cite{Ammar2014, Yassin2016} the authors reviewed various technologies for indoor localization and assessed the performance of each indoor localization technique. However, localization is not discussed in terms of energy efficiency or any prospective application. Additionally, the authors did not explore different techniques to enhance the localization accuracy. In \cite{Vo2016} a remarkable survey is presented on fingerprinting-based localization systems. Recently, in \cite{Ferreira2017}, the authors have presented a survey on indoor positioning system mainly focusing on the emergency applications. In \cite{Tian2012} the authors have presented different possible architectures for MDS based localization for WSNs. However, the paper is focused only on the different variants of MDS schemes and does not cover all the aspects of MDS based localization schemes. The aim of this survey is to present a comprehensive overview of localization systems to cover both outdoor and indoor localization systems with the main focus on the development of MDS based localization schemes from its inception to its current state for different applications.

\subsection{Survey Organization}
The remainder of this survey article is organized as follows. In Section \ref{sec:global} we present a detailed survey on different outdoor and indoor positioning systems. Section \ref{sec:ranging} introduces the fundamentals of different ranging techniques. In Section \ref{sec:MDSbasics}, we focus on MDS technique and cover different variant of MDS based localization methods. Section \ref{sec:MDSlocalization} covers the literature on different MDS based localization methods used for various wireless networks. Section \ref{sec:applications} covers the prospective applications of MDS based localization.  Section \ref{sec:conc} summarizes the survey paper and conclude the work. 

To summarize the different features of this survey which differentiate it from the existing works; first, a brief review of advanced positioning systems for outdoor and indoor environments is presented. Second, we discuss the different ranging techniques used by localization systems. Third, technical details of MDS techniques are covered along with its usage for localization systems. Fourth, we review different localization systems for various wireless networks based on the MDS method. Finally, we compare the MDS based localization schemes and present applications of MDS based localization method.

\section{Overview of Positioning Systems and Ranging Techniques}\label{sec:global}

\begin{figure*}[htb!]
\begin{center}  
\includegraphics[width=2\columnwidth]{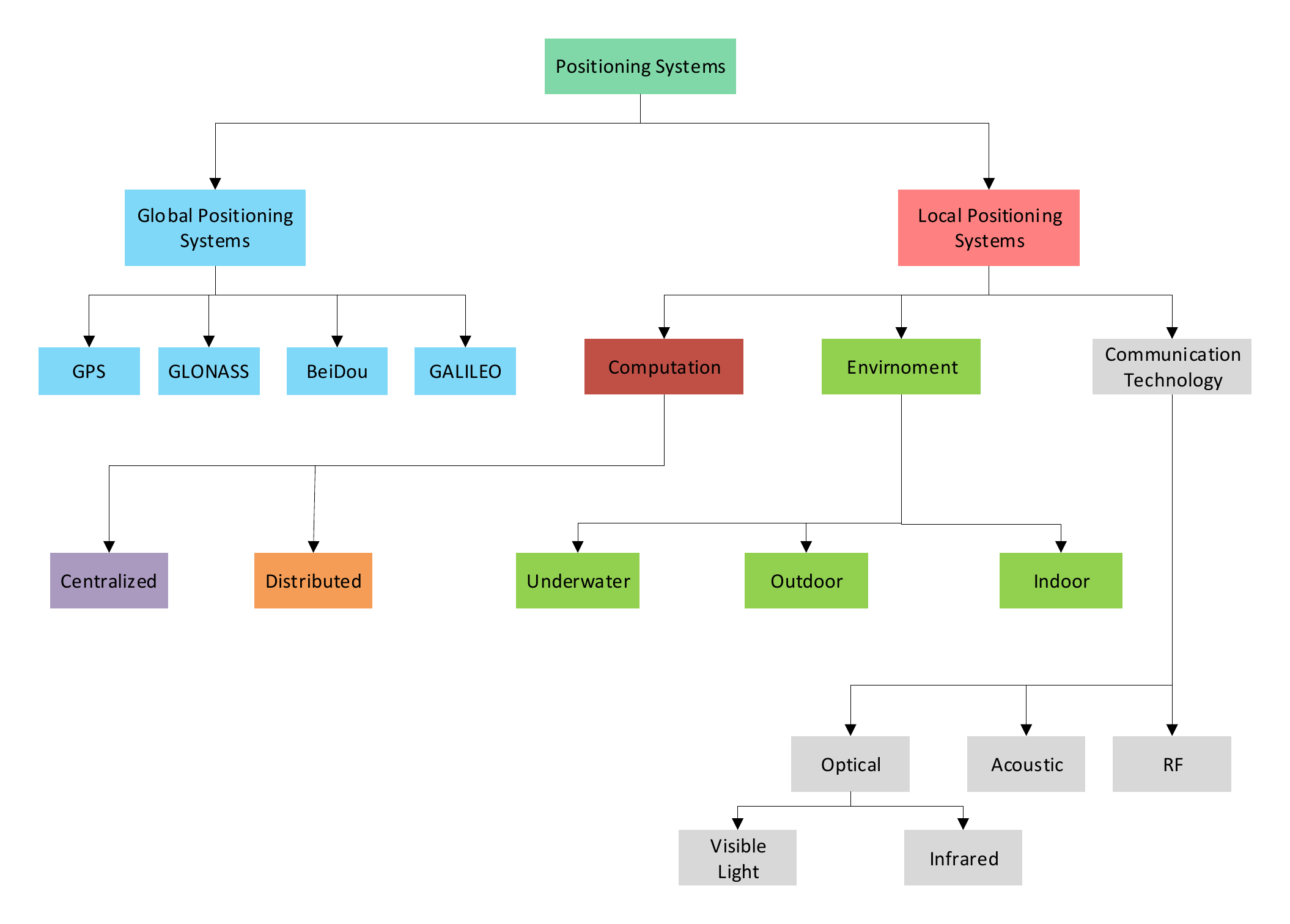}  
\caption{Classification of positioning systems.\label{fig:classification}}  
\end{center}  
\end{figure*}

In this section, a brief overview of global and local positioning systems (LPS)  is presented. Positioning systems are broadly categorized into two major categories global positioning systems and local positioning systems as shown in Fig.~\ref{fig:classification}. Moreover, various ranging techniques used for localization systems are also discussed.

\subsection{Global Positioning Systems}
Global positioning systems are the systems that use satellites to provide location information to the user. Global positioning systems allow the users to determine their locations with the accuracy of a few meters in the outdoor environment. The global coverage can be achieved with the help of multiple global positioning systems such as GPS, GLONASS, GALILEO, and BeiDu.
\subsubsection{Global Positioning System (GPS)}
The global positioning system (GPS) is one of the most common and successful positioning systems in outdoor environments, which consists of 28 operational earth orbiting satellites. A user or an object with a GPS receiver can localize itself in terms of longitude, latitude, and altitude with the accuracy of a few meters \cite{Dougherty}. Satellites orbit around the earth at the height of 12,000 miles and accomplish two rotations every 24 hours. The particular characteristics of the GPS satellites are such that at any time anywhere on the earth surface at least four satellites are visible \cite{Reza2011}. 
The concept of GPS based localization requires precise time and the position of the satellites. Highly stable atomic clocks are carried by the satellites which are synchronized with the clocks on the ground segment as well as with each other. Similarly, the locations of satellites are known with high precision. The GPS receivers clocks are cheap, less stable and not synchronized with the satellite clock. GPS satellites continuously broadcast its time and location and the receiver computes the pseudo-ranges from each visible satellite. The receiver needs to have at least four satellites visible at the time of calculating the four unknowns (three location coordinates and a clock offset).
\subsubsection{Global Navigation Satellite System (GLONASS)}
Global Navigation Satellite System (GLONASS) is a space-based navigational and localization system operated by Russia, which provides an alternative to the GPS \cite{Dale}. GLONASS does not have broad coverage like GPS, yet the coverage and accuracy are certainly increased when both GPS and GLONASS are used together. GLONASS has an accuracy of up to 2 meters. The use of GPS with GLONASS allows users to be precisely positioned by a league of 55 satellites covering the globe. Therefore, when a user is in a location where GPS signals are blocked in an urban area by huge buildings, a user can be located by GLONASS satellites. A lot of more smart-phones are being introduced with GPS+GLONASS technology to provide location-based services. For example, various localization, and tracking products of Wialon use both GPS and GLONASS signals \cite{Glonass2}. Similarly, the integration of GPS and GLONASS signals are studied in \cite{Przestrzelski2017} for improved coverage.
\subsubsection{GALILEO System}
GALILEO is another navigational and positioning system owned by the European space agency, delivering an extremely accurate, reliable global positioning facility under civilian control. GALILEO has inter-operability with GPS and GLONASS \cite{Kacmarik}. GALILEO system consists of 27 active and 3 spare satellites circulating the earth at an altitude of 24000 km. GALILEO and GPS have a similar bandwidth and center frequency band which means that GALILEO system is smoothly interoperable with GPS and its signal performance is far better than GPS \cite{Fernandez}. The signal performance enhancement of GALILEO system is due to the use of novel modulation technique called composite binary offset carrier (CBOC) which improves the received power almost to the double of the C/A coded GPS signals \cite{Galileo2}. Also, in spite of bringing in some more frequency band signals, the GALILEO system introduces a very little complexity to the receiver design \cite{Hui1}.
\subsubsection{BeiDou Navigation Satellite System (BDS)}
BeiDou Navigation Satellite System (BDS) is a Chinese satellite navigation system also called COMPASS. It consists of 12 operational satellites including five geosynchronous satellites, four medium earth orbit satellites, and three inclined geosynchronous orbits satellites \cite{Dardari2011}.  BDS became operational in China in December 2011 \cite{Canhui, Grelier} and began services in December 2012 in the Asia-Pacific region. Last year, 19  more satellites were launched in several orbits providing the accuracy up to 10 meters globally and up to 5 meters in the Asia Pacific region. BDS system provides high accuracy and reliability, support inter-satellite links, and augmentation systems \cite{beiduo2}.

\subsection{Local Positioning Systems}

LPS provide location information to the user with the help of base stations or anchors which can generate beacon signals. The coverage of LPS is limited, and localization is achieved only within the coverage area of the network. LPS can be categorized based on different network criteria, here we broadly classify them by availability of computation, environment, and medium for transmission. 

\subsubsection{Computation}
LPS can be broadly categorized into distributed and centralized techniques, based on the computation \cite{Reza2011}.
In distributed positioning systems every user can determine its location with the help of geographically distributed anchors. Many distributed positioning systems have been presented in the past for WSNs such as \cite{Mondinelli2004} and \cite{Stein2017}. Unlike distribute LPS, in centralized positioning systems, each user determines its neighborhood information using time of arrival (ToA), angle of arrival (AoA), time difference of arrival (TDoA), and received signal strength (RSS). The neighborhood information is collected at a centralized station which finds out the location of the user and shares the location information with the user.

\subsubsection{Environment} Since every positioning system heavily depends on the environment, different LPS have been developed for different environments. These LPS can be divided into three categories based on the environment, i.e., outdoor LPS, indoor LPS and underwater LPS. 
\begin{itemize}
\item Outdoor LPS: Localization in outdoor is usually provided by GPS with an accuracy of 5 to 10 meters. With the help of wide area augmentation systems, the accuracy is improved to the range of 1 to 8 meters. But still this accuracy is not sufficient for certain applications, therefore, NavCom provided a local differential GPS based outdoor positioning system with an accuracy of 1 centimeter. However, the power constraint and higher cost generally do not allow the use of GPS receiver for small sensor devices. Therefore, GPS less outdoor positioning was proposed in\cite{Bulusu}. In \cite{ Agroudy2016} the authors proposed a low power consumption localization scheme for outdoor positioning by using a power management scheme. Unfortunately, till date, the academic proposals \cite{Cheng2005, Tsu2010} as well as the industrial practices \cite{rcell, skyhook} for outdoor LPS have not achieved satisfactory localization accuracy.

\item Indoor LPS: In recent years indoor LPS has attracted great attention due to its commercial and social values, where the predicted market value for indoor LPS worth 10 billion US dollars by 2020. Indoor environments are more complex which is characterized by a large number of obstacles, signal fluctuations, noise, environmental changes, and non-line of sight communication. Despite such complexity, accurate indoor LPS are required for satisfactory indoor LBS.
Majority of the research efforts have been made in the past two decades to develop accurate, low cost, and energy efficient indoor LPS. For more details on indoor LPS,  interested readers are referred to the survey articles presented on this subject such as  \cite{Yassin2016}, \cite{ Adler2015}, and \cite{CHOWDHURY2016}.

\item Underwater LPS: A number of underwater LPS have been proposed in the past for underwater acoustic wireless communication systems. All of these localization algorithms consider different parameters of the network such as network topology, range measurement technique, energy requirement, and device capabilities.  In addition, the accuracy of localization algorithms also depends on many other factors which include propagation losses, number of anchor nodes, the location of anchor nodes, time synchronization, and scheduling \cite{Ramezani2015}. Thus, many researchers developed localization schemes which take into account the above factors for acoustic waves-based underwater localization. Hence, a few brief surveys are presented on this subject such as \cite{kantarci2011survey} and \cite{Tuna2017}. However, the speed of acoustic waves is slow and therefore it leads to the development of high speed underwater optical wireless communication (UOWC) systems. In comparison with the acoustic systems, UOWC can support higher data rates  up to several Gbps in clear waters with little to no scattering. However, UOWC suffers from low transmission range and require accurate pointing between the transmitter and the receiver. To provide the localization capabilities in UOWC systems, various localization schemes have recently been developed such as \cite{Saeed2017} and\cite{Nasir2018limited}.

\end{itemize}
\subsubsection{Transmission Medium}
LPS can also be categorized based on the transmission medium.
\begin{itemize}
\item Radio Frequency (RF) based LPS: The popular LPS based on RF technology consists of cellular networks based LPS, wireless local area networks (WLANs) based LPS and radio frequency identification (RFID) based LPS.
\begin{itemize}
\item Cellular networks based LPS: LPS based on cellular networks has been discussed for more than a decade. Initially, the position of a mobile terminal was determined using global system for mobile communication (GSM) \cite{Drane1998}. Indeed the techniques discussed in \cite{Kyamakya2005} influenced the standardization of universal mobile telecommunications system (UMTS). In cellular network based LPS, the location of the mobile station is determined by the base station by using the cell geometries. 
Interested readers are referred to \cite{ Peral2017}, which is the most recent survey article on cellular-based LPS.

\item WLANs based LPS: WLAN-based LPS are very popular among other LPS due to its established infrastructure.  In \cite{Bahl} the authors proposed a LPS which can locate and track the user using the nearest neighbors technique. The accuracy of this WLAN based LPS is 2 to 3 meters. 
There are several other WLAN based LPS, for the interested readers  we refer to the detailed surveys presented in \cite{Nuaimi2011,  Liu2007, Honkavirta2009, KUL2014, Khalajmehrabadi2017} on this subject.

\item RFID based LPS: RFID technology is mostly employed in harsh indoor scenarios such as offices, hospitals, subways etc. RFID based technology provides cheap and adaptable identification of a device or an individual\cite{Chon}. 
For supporting indoor and outdoor localization in real time, WhereNet is the popular real-time location system (RTLS) offered by Zebra technology \cite{Wherenet} which is based on RFID tags and differential time of arrival (DToA) technique.

\end{itemize}

\item Acoustic based LPS: Acoustic waves are also used in localization systems to locate a node or a user \cite{Lazik2015, Ward, Woodman, Priyantha1, Holm, Dolphin}. It is known that bats use acoustic signals to navigate. Inspired by this, Active Bat localization system was developed by AT \& T  based on acoustic signals, which provides 3-dimensional localization. Active Bat localization system consists of an acoustic system and triangulation approach for localization. The distance between the transmitter and receiver is measured through ToA measurements. Some other major acoustic based LPS are Cricket \cite{Priyantha1}, Sonitor \cite{Holm}, and DOLPHIN \cite{Dolphin}.

\item Optical LPS: Optical LPS are becoming dominant LPS  which covers a wide range of applications. Optical LPS can further be classified into visible light communications (VLC) based LPS and infrared-based LPS.

\begin{itemize}
\item Visible Light based LPS: The advancement of visible light technology has led to the development of visible light based communication (VLC). Based on the universality and recent research on VLC, LPS are considered to be an important feature of VLC. A theoretical accuracy of centimeters has been reported in \cite{Cai2017, Steendam2016, Lv2017} by using VLC for LPS. Recently, a number of practical LPS such as Luxapose \cite{Kuo2014}, PIXEL \cite{Yang2015}, Epsilon \cite{Li2014}, and LIPS \cite{Xie2016} based on VLC are proposed. Epsilon was the first visible light based LPS which can achieve an accuracy of 0.4 to 0.8 meters. Luxapose, LIPS, and PIXEL achieve an accuracy of 0.1, 0.4, and 0.2 meters respectively. It should be noted that every LPS based on visible light strongly depends on the light emitting diode (LED) technology, types of receivers, and modulation method used. Interested readers are referred to \cite{Luo2017} and \cite{saeed2018camera} where the authors have reviewed a number of LPS based on VLC.

\item Infrared (IR) based LPS: Infrared (IR) based LPS \cite{Casas, Want, Harter, Firefly, States, Aitenbichler} are the most common localization systems owing to the availability of the IR technology for numerous gadgets. IR based localization system requires line of sight (LOS) connection between the transmitter and receiver in the absence of any kind of interference. Some of commercial IR based localization and tracking systems are Firefly \cite{Firefly}, OPTOTRAK \cite{States}, and infrared indoor source local positioning system (IRIS\_LPS) \cite{Aitenbichler}.
\end{itemize}

Major issues with optical LPS include multipath reflections, synchronization, coverage, and privacy. For example, optical LPS require line of sight (LoS) links for range estimation. However, LoS link may not always be available due to the multi-path effect caused by light reflections from various surfaces. Similarly, synchronization is also a significant issue for time-based ranging in optical LPS because it is challenging to synchronize all the transmitters and the receivers. Limited coverage of the LED transmitters is also of major concern due to their directive nature, for optical LPS.
\end{itemize}

\subsection{Fundamental Ranging Schemes}\label{sec:ranging}
The fundamentals ranging techniques used for range based localization systems are discussed in this sub-section. All of the above positioning systems depend on the ranging measurements. Following are the different ranging schemes for distance estimation.

\subsubsection{Time of arrival (ToA) Estimation}
ToA is one of the most widely used ranging techniques for positioning systems. In ToA based positioning systems, users computes the time delay of signal propagation to estimate the distance between the receiver and the transmitter. The main problem with ToA measurements is that the received signal arrives through multipath with different delay through the channel \cite{Guvenc}. LoS signal is presumed to be available in ToA systems to compute the signal propagation delay \cite{Gu}.
In ToA systems setup, the anchors broadcast the beacon signal while the node (or user) computes propagation delay of the received signal from multiple anchors. The transmitted signal travels with the speed of light and thus, the distance between node and anchor is estimated from the propagation delay. The intersection of the circles from different anchors leads to the region of estimated position of a node. But due to different environmental effects, the signal arrives at the node at multiple paths with different delays \cite{Karalar2004}. Therefore, multipath leads to an error in position estimation, because the circles from different anchors do not intersect at a single point \cite{Gu}.


\subsubsection{Time Difference of Arrival (TDoA) Estimation} 
TDoA is an enhanced version of ToA technique where a node estimates the distance by receiving two different kinds of signals from the same anchor or same type of signal from two different anchors.  Cricket system \cite{Priyantha1} is a good example of TDoA based indoor localization system which uses ultrasound signals and RF signals for localization. The time difference between the two signals is calculated by the receiver and generates a hyperbola  \cite{Rappaport}. The point where the hyperbolas from different anchors intersect yields to the node position. In comparison to the ToA measurements, the TDoA does not require synchronization between the anchors and the node \cite{noureddinealmoussawi2012}. But using two different kinds of signals for localization leads to a higher cost due to the extra hardware required to transmit and receive two different kinds of signals.

\subsubsection{Received Signal Strength (RSS) Estimation}
Received signal strength (RSS) measurements are the simplest and most commonly used technique for distance estimation \cite{Moravek2013}. The free space path loss model is usually used to estimate the distance from the measured received power \cite{rss}. The strength of a received signal is decreased due to path loss, frequency selective fading, and shadowing. The effect of path loss is to be measured, as it is a deterministic decrease in power as a function of the distance between the node and the anchor. Multipath fading, in spite of being problematic, is deemed advantageous. This fading is produced by either constructive or destructive addition of time-delayed signals at various frequencies. Therefore, the correlation between the estimations is less if the estimations are carried out at different frequencies that are separated beyond the coherence bandwidth. Furthermore, multiple spread-spectrum wireless sensors will be employed that will average out frequency selective fading. Unfavorably, the same technique is not present to counter shadowing, which is most of the times introduced by object blockage between the node and the anchors. 
The received power $P_r(d)$ at distance $d$ can be written as \cite{Jin2003}
\begin{equation}\label{path loss}
P_r(d) = P_r(d_0)-10\eta\log\frac{d}{d_0},
\end{equation}
where $P_r(d_0)$ represents the power received at reference distance $d_0$ and $\eta$ is the path loss exponent. 

%
\begin{table*}
\centering
\caption{Devopment of  MDS\label{table2}}{
\begin{tabular}{|c| r|r|} 
\hline\hline 
\bf MDS methods & \bf Authors & \bf Year  \\ [0.5ex] 
\hline
Foundation of MDS & Eckart and Young & 1936-1938  \\
Classical MDS   & Torgerson & 1952 \\
Principal component analysis & Gower & 1966 \\
Non-metric MDS & Shepard and Kruskal & 1962-1964 \\
\hline
\end{tabular}}
\end{table*}
\subsubsection{Fingerprinting}
Fingerprinting approach is based on the fact that radio waves emitted from the base stations leave a unique radio fingerprint at a given location that can be used for localization \cite{Olivier2016}. The radio fingerprint is obtained by creating a database of the average values of RSS from various anchors at different locations. This requires a training phase to collect the fingerprints at known locations which can be used for the localization of the user based on probabilistic or deterministic positioning techniques, e.g., maximum likelihood estimator or $k$-nearest-neighbor estimator. 
Presently, most of the indoor localization methods are based on fingerprint matching technology \cite{finger3}. Researchers have employed different methods to make fingerprint matching technology better in all aspects. As compared to other localization systems, Wi-Fi fingerprint positioning technology is cheap and has great precision. Owing to the vast deployment and use of Wi-Fi all over the world, fingerprint positioning technology can be used in any indoor environment where Wi-Fi networks are established, without the installation of extra hardware. In a complicated indoor scenario, under harsh conditions, the space-time traits such as angle and time of arrival can be erroneous, but the signal intensity is relatively stable. Therefore, it makes the accuracy of fingerprinting-based localization higher than other techniques. 
Fingerprinting based positioning systems are reviewed comprehensively in \cite{Khalajmehrabadi2017, Vo2016, BASRI2016}.

\subsubsection{Direction of Arrival (DoA) Technique}
DoA ranging measurements are based on the angle of the received signal at the receiver \cite{aoa}. The DoA-based approaches are simpler than time-based techniques because only two angle measurements are required to estimate the two-dimensional position. However, obtaining the accurate DoA-based ranges is a challenging task, especially in NLoS conditions. Moreover, in the indoor environments where the LoS signal is hard to obtain, DoA measurements are highly erroneous. 
DoA based techniques are classified into the following two categories based on the applications:
\begin{itemize}
\item Online DoA: These techniques have lower complexity and are used for applications which require real time location information. In online DoA method, the angles are determined from the received signals and by using geometrical relationship (tri-angulation) between the anchors' position and the source position, the location of the source is estimated. 
\item Offline DoA: These techniques have high complexity and can only be used for offline applications. Offline DoA is similar to the fingerprinting technique, where the DoA measurements are calculated multiple times and the average value is designated as the fingerprint. The source then locates itself by using these fingerprints by using triangulation. 
\end{itemize}
Online DoA is used in applications where high precision is not important such as beam-forming and signal detection \cite{Godara1996}. Localization applications need accurate DoA estimation, even if it is not online. In comparison to other ranging techniques, DoA is more accurate, but consume high power and have greater complexity \cite{Kumar}.

\begin{table*}
\centering
\caption{Development of loss function for MDS\label{table3}}{
\begin{tabular}{|c| |r|r|}  
\hline\hline 
\bf Loss functions & \bf Authors & \bf Year  \\ [0.5ex] 
\hline
Sammon’s  mapping & Sammon & 1969  \\
Coombs Unfolding model  & Coombs & 1964 \\
Carroll models & Carroll and Chang & 1970 \\
ALSCAL & Takane & 1977 \\
Maximum likelihood& Ramsay & 1982\\
Optimal scaling& Meulman & 1992-1993\\
\hline
\end{tabular}}
\end{table*}

\section{Multidimensional Scaling Based Localization}\label{sec:MDSbasics}
This section briefly introduces the basics of MDS and review a number of MDS based localization techniques for applications in WSNs-IoT, cognitive radio networks, and 5G networks.
\subsection{What is MDS?}
MDS is a dimensionality reduction method which converts a higher dimensional data into a lower dimension. Due to this dimensionality reduction provided by MDS, it can display the data graphically which is more meaningful and easy to understand. There is a large number of dimensionality reduction methods such as factor analysis, principal component analysis, and Isomap. But due to the simplicity and wide range of applications, MDS is most popular. 

The input for any MDS based method is a dissimilarity or similarity information among the objects or points \cite{Borg,Bronstein,Joseph}. The MDS method uses this dissimilarity or similarity information and tries to closely match it to the Euclidian distance between those objects or points\cite{Honar,Cambria,bai2016multidimensional,gansner2004graph}. Unlike factor analysis, MDS does not depend on the assumptions of linearity and normality \cite{Biswas}. The only assumption required for MDS is that the number of dimensions required should be one less than the number of points \cite{Bronstein1}.

Since MDS is one of the classical data analysis methods used in wide range of applications, therefore, rich literature exists on MDS methods for achieving data visualization and data analysis \cite{borg2012applied,WCS:WCS1203,Saeedcomput}. Results on classical MDS and its recent variants are briefly discussed in \cite{T.Cox, Borg}. The MDS method was originated by Eckart and Young \cite{ Eckart, Young2}, while the first input metric for MDS was developed by Torgerson \cite{Torger}. In \cite{Gower, collins} the authors established a relationship between MDS and principal component analysis. The non-metric MDS was developed by Shepard and Kruskal in \cite{Kruskal1} where the dissimilarity or similarity information relates monotonically to the Euclidian distances \cite{Nadia}. Table \ref{table2} shows the development of MDS methods over the years. 

\begin{figure}[htb!]
\begin{center}  
\includegraphics[width=1\columnwidth]{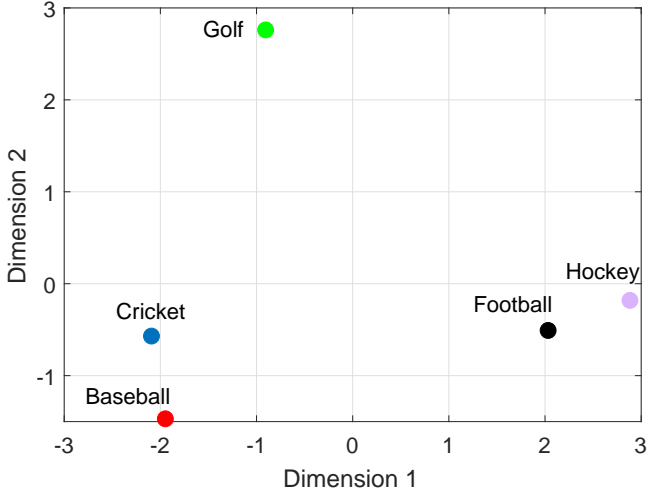}  
\caption{Example of MDS map for various sports visualization.\label{fig:sports}}  
\end{center}  
\end{figure}

Every MDS method can be specified by its loss function \cite{Sammon} and a number of different loss functions have been developed for MDS methods. Some of the famous loss functions for MDS are Coombs unfolding model \cite{Coombs}, individual difference model \cite{Carroll1}, ALSCAL \cite{takane}, maximum likelihood \cite{Ramsay}, and optimal scaling \cite{Meulman1}. Table \ref{table3} summarizes different loss models used for MDS methods. The loss functions basically relates the measured values (dissimilarities)  to their Euclidean distances. To elaborate more, consider that the dissimilarity between any two points $i$ and $j$ is $\rho_{ij}$ and their corresponding Euclidean distance is $d_{ij}$, then the squared error function is represented as
\begin{equation}
e^2_{ij} = \left(\rho_{ij} - d_{ij}\right)^2.
\end{equation}
Based on the squared error function, the total error (raw Stress) for all pair of objects is obtained as
\begin{equation}
e_r = \sum_{i=1}^n\sum_{j<i}^n\left(\rho_{ij} - d_{ij}\right)^2,
\end{equation} 
where $n$ is the total number of objects. The major problem with the above raw stress function is that it is invariant under coordinate scaling transformation. Hence, normalization techniques are used to address the problem of in-variance. One of the most proper choice for the normalization is using the dissimilarity, i.e., $\rho_{ij}$. By using $\rho_{ij}$ as a normalization parameter yields the well-known loss function for MDS called Kruskal stress function which is given as
\begin{equation}
e_s =\sqrt{\frac{\sum_{i=1}^n\sum_{j<i}^n\left(\rho_{ij} - d_{ij}\right)^2}{\sum_{i=1}^n\sum_{j<i}^n\rho_{ij}^2}}.
\end{equation}
This loss function can be solved by using the well-known iterative majorization approach called scaling by majorizing of a complicated function (SMACOF). To further elaborate, we consider a simple example of various sports classification. The range of dissimilarities is set to $1 = \text{very similar}$, $3= \text{average similarity}$, and $5= \text{non-similar}$, respectively. We consider five different sports which include cricket, baseball, hockey, football, and golf. The dissimilarity matrix between these sports is given in Table. \ref{Sports}. Applying classical MDS to this  dissimilarity matrix yields a graphical map (Fig. \ref{fig:sports}) of these sports which shows the relationship between these sports in a two-dimensional space. For example, Fig. \ref{fig:sports} shows graphically that cricket and baseball are similar sports, similarly, hockey and football are similar while golf is different than all the other sports.  Note that this MDS map has no real orientation  which means that it can be rotated around its center. The main characteristics are the relative positions of each point.
\begin{table}[]
\caption{Dissimilarity matrix.} \label{Sports}
\begin{tabular}{|l|l|l|l|l|l|}
\hline
Sports   & Cricket & Baseball & Hockey & Football & Golf \\ \hline
Cricket  & 0                               & 1                                & 5                              & 5                                & 3                            \\ \hline
Baseball & 1                               & 0                                & 5                              & 5                                & 5                            \\ \hline
Hockey   & 5                               & 5                                & 0                              & 1                                & 5                            \\ \hline
Football & 5                               & 5                                & 1                              & 0                                & 5                            \\ \hline
Golf     & 3                               & 5                                & 5                              & 5                                & 0                            \\ \hline
\end{tabular}
\end{table}

Traditional MDS methods consider that distances among objects are symmetric, although this consideration is not always satisfied. For instance, \cite{tversky1977features} and \cite{tversky1978studies} explained the characters of similarity among objects studied with psychological scale and concluded that cognitive similarity is mostly asymmetric.
The motivation behind these asymmetric MDS methods is to remove the shortcomings of traditional MDS methods, i.e., in case where similarity or dissimilarity matrices are asymmetric in nature as they are based on the supposition that similarity or dissimilarity matrices can be associated with inter-point distances in a given metric space \cite{torgerson1952multidimensional}, \cite{kruskal1964multidimensional}, \cite{guttman1968general}. Many researchers have extended the traditional MDS methods by assuming that the similarity or dissimilarity among objects is not a function of only inter-point distances but is also a function of the quantities associated with these objects. For instance, the squared distances are extended by weights in weighted distance model which was first proposed by \cite{young1975asymmetric}. In \cite{saito1983method}, \cite{saito1986multidimensional} and \cite{weeks1982restricted} the authors have proposed altered distance models where the distance between points is established by a few constants associated with these points. In \cite{okada1987geometric} and \cite{okada1987nonmetric} a nonmetric type of generalized altered distance model is proposed. Smallest space analysis-2 (SSA-2) is introduced  in \cite{guttman1968general} and \cite{lingoes1973guttman}, where column and row compatibility is applied on the data to get two solutions in metric space. Wind model is proposed in \cite{tobler1975spatial}, where the asymmetries are explained by the direction of wind given to mesh point on the arrangement of objects. In \cite{saito1988cluster} and \cite{sato1989distance} the authors proposed a model, in which the asymmetries are analyzed by utilizing the Randers metric i.e., an asymmetric metric function.  In \cite{tversky1977features} the authors proposed the feature matching model which explains the similarity or dissimilarity among two objects through a linear combination of the amount of distinctive and common characteristic of the two objects. In
\cite{chino1978graphical} and \cite{chino1980unified} the authors proposed a model using a generalization of scalar products, which fits the magnitude of cross and inner (scalar) products of solution vectors to skew-symmetric and symmetric parts of the data, respectively. 
\cite{constantine1978graphical} and \cite{gower1977analysis}, split the asymmetric proximity matrix into two components, i.e., symmetric and skew-symmetric components and then deal with them separately. For symmetric component, traditional MDS method is used, while for skew-symmetric component canonical decomposition is used. In \cite{saburi2008maximum} the authors proposed a maximum likelihood method for asymmetric proximity matrix, which expands the work for asymmetrical data \cite{takane1981multidimensional}.  

MDS maps the original high dimensional data ($m$ dimensions) in to a lower dimensional data ($d$ dimensions). It addresses the problem of constructing a configuration between the $n$ points from $n\times n$ matrix $\boldsymbol{D}$, which is called  distance affinity matrix and it is symmetric, i.e., $d_{ii}=0$, and $d_{ij}>0,~i\neq j$. MDS finds $n$ data points $\boldsymbol{P}=\{\boldsymbol{p}_{i}=\{x_i,y_i\},...,\boldsymbol{p}_{n}=\{x_n,y_n\}\}$ from the distance matrix $\boldsymbol{D}$ in a $d$ dimensional space, such that the estimated distance $\hat{d}_{ij}$  between $\boldsymbol{p}_{i}$ and $\boldsymbol{p}_{j}$, matches the Eucleadian distance as closely as possible. In \cite{ghodsi2006, cox2008multidimensional}, the loss function for MDS is  considered as 
\begin{equation}\label{eq:loss}
L(\boldsymbol{P})=\sum_{i=1}^{n}\sum_{j< i}^{n}{(\hat{d}_{ij}-d_{ij})}^{2},
\end{equation}
which is highly nonlinear. To solve it, the distance affinity matrix $\boldsymbol{D}$ is first converted to a kernel matrix of inner product $\boldsymbol{P}^{T}\boldsymbol{P}$ by
\begin{equation}\label{eq:loss2}
\boldsymbol{P}^{T}\boldsymbol{P}=-\frac{1}{2}\boldsymbol{H}\boldsymbol{D}\boldsymbol{H},
\end{equation}
where $\boldsymbol{H}=\boldsymbol{I}-\frac{1}{n}\boldsymbol{ee}^{T}$ is called the double centering operation, $\boldsymbol{I}$ is identity matrix of size $n \times n$, and $\boldsymbol{e}$ is a column vector of $1$'s. 
The solution of \eqref{eq:loss2} is $\boldsymbol{Y}=\Lambda^{1/2}\boldsymbol{V}^{T}$ where $\boldsymbol{V}$ are the eigen-vectors of $\boldsymbol{P}^{T}\boldsymbol{P}$ in $d$ dimensions and $\Lambda$ are the $d$ eigenvalues of $\boldsymbol{P}^{T}\boldsymbol{P}$. 

In recent past MDS is widely is used for simultaneous localization and mapping of WSNs and IoT networks.  In \cite{Shang} a proximity information based sensor network localization is proposed. The main idea in \cite{Shang} is to construct a local configuration of sensor nodes using classical MDS (CMDS). The analogy between object distances and node distances in a network is used for the purpose of WSNs-IoT localization. MDS algorithm uses inter-node distances in order to produce two or three-dimensional representation, which corresponds to the real nodes deployment. Since nodes are capable to measure the inter-node distances with respect to their neighboring nodes, the only problem remains to obtain the non-neighboring inter-node distances. In MDS method, these distances are approximated by using Floyd Warshall shortest path algorithm \cite{Floyd}.

Distances between every node in the network are collected at the central station. The remaining (non-neighboring) distances are calculated by the central station.  The calculation for 2D network consists of the following steps:

\begin{itemize}
\item Compute the shortest path distances between every node in the network (using either Dijkstra or Floyd algorithm). These shortest path distances work as an input data for MDS.

\item Classical MDS is applied to the shortest path distance matrix which results in the spectral decomposition of input data matrix. The two largest eigenvalues and the corresponding eigen-vectors form the relative location of every node in the network (three largest Eigenvalues and eigen-vectors for 3D localization).

\item Finally, the relative locations are transformed to the absolute global locations by using the anchor nodes. This transformation includes optimal rotation, translation, and reflection. This type of transformation is also called rigid or Euclidean transformation \cite{Eggert}.
\end{itemize}
Based on the computation, MDS based localization methods have been proposed in the past which can be categorized into centralized, semi-centralized, and distributed methods. 

\subsection{Centralized MDS Based Localization}\label{centralizedsec}
Assume that there are $n$ nodes in the network and the pairwise range measurements between the nodes are noisy, the centralized MDS based localization consists of the following steps \cite{T.Cox}
\begin{itemize}
\item The shortest path distances are estimated between each pair of all nodes by using shortest path algorithms (Dijkstra or Floyd Warshall algorithms \cite{Floyd}) to construct the distance affinity matrix $\boldsymbol{D}={\{\hat{d}_{ij}^2\}^n_{i,j=1}}$ which can be written in matrix form as
\begin{equation}\label{eq: d1}
\boldsymbol{D}=\begin{bmatrix}
\hat{d}^2_{11}& \cdots & \hat{d}^2_{1n} \\ \vdots & \ddots & \vdots \\ \hat{d}^2_{n1} & \cdots & \hat{d}^2_{nn} \end{bmatrix},
\end{equation}
where the direct neighborhood distance is $\hat{d}_{ij} = d_{ij}+\epsilon_{ij}$.  $d_{ij}= \sqrt{{({x}_i-x_j)^2+({y}_i-y_j)^2}}$ is the Euclidean distance between node \textit{i} and \textit{j},  
$\epsilon_{ij}$ represents the ranging error which is modeled as zero-mean Gaussian random variable with variance $d_{ij}\eta_{ij}^2$ where $\eta_{ij}^2={\mu} {d_{ij}^{\beta_{ij}-1}}$, $\mu$ is scalar constant related to the receiver, and $\beta_{ij}$ is the path loss exponent. As the matrix $\boldsymbol{D}$ is square symmetric with $\hat{d}_{ii}=0$ and $\hat{d}_{ij}=\hat{d}_{ji}$, therefore, simplifying \eqref{eq: d1} yields
\begin{equation}\label{eq: d2}
\boldsymbol{D}=\begin{bmatrix}
0 & \cdots & \hat{d}^2_{1n} \\ \vdots & \ddots & \vdots \\ \hat{d}^2_{n1} & \cdots & 0 \end{bmatrix}.
\end{equation}
\item The MDS method is applied to the distance affinity matrix $\boldsymbol{D}$ to minimize the discrepancies between the actual Euclidean distances and estimated distances. The normalized loss function or stress function for the MDS method is defined as 
\begin{equation}\label{eq: stress}
S(\hat{d}_{ij}|\boldsymbol{P}) = \frac{\sqrt{\sum_{i\neq j=1...n}\left(\hat{d}_{ij}-d_{ij}\right)^2}}{\sum_{i\neq j=1...n}(\hat{d}_{ij})^2}.
\end{equation}
The stress function defined in \eqref{eq: stress} is nonlinear and nonconvex, therefore to get the close form solution for this function, the distance affinity matrix $\boldsymbol{D}$ is double centered by using the double centering operator $\boldsymbol{H}=\boldsymbol{I}-\frac{1}{n}\boldsymbol{ee}^{T}$, given as
\begin{equation}\label{eq: doublecenter}
\boldsymbol{C}=-1/2(\boldsymbol{H}\boldsymbol{D}\boldsymbol{H}),
\end{equation}
which is then decomposed by using Eigen value decomposition given as
\begin{equation}\label{eq: doublecenter2}
\boldsymbol{C}= \mathbf{e}\boldsymbol{\lambda}\mathbf{e}^T,
\end{equation}
where $\mathbf{e}$ represents the eigen-vectors and $\boldsymbol{\lambda}$ are the eigenvalues.
Finally, the relative two dimensional positions of the nodes are determined from the two largest eigenvalues of $\boldsymbol{\lambda} $ and two largest eigen-vectors in $\mathbf{e}$ i.e.,
\begin{equation}\label{eq: rel1}
\mathbf{\hat{P}}= {\mathbf{e}_2\sqrt{\boldsymbol{\lambda}_2}},
\end{equation}
Since the position estimates obtained in \eqref{eq: rel1} are not absolute, it is required to transform these relative position estimates into absolute (global) positions. Linear transformations such as Procrustes analysis, Helmert transformation, or principal coordinate analysis can be used to get the global position estimates.
\end{itemize}
Centralized MDS based localization method was first proposed by Shang in \cite{Shang}. The proposed method in \cite{Shang} is applicable to both range based and range free conditions. The benefit of using centralized MDS is that it can work with few number of anchors with high accuracy. But the problems with this approach are high computational overhead and large localization error for irregular networks. An ordinal MDS based centralized localization method is proposed in \cite{viv} which requires only the relationship between the shortest path distances and the Euclidean distances. Classical centralized MDS based localization is proposed in \cite{Shi2011} for RFID systems. Centralized RSS based non-metric MDS is used in \cite{Gao2017} to find the location of RFID tags. Small scale WSNs localization is investigated in \cite{Weber2011, Chun2015} by using centralized MDS method. A centralized cooperative MDS based localization method is proposed in \cite{Chen2016} for WLANs. Authors in \cite{Wang20015} have investigated the multipath propagation ranging error for MDS based localization method. In \cite{Bingjie2010} a weighted MDS is proposed for WSNs localization where the accurate range measurements are given more weight, and the noisy ranges are down-weighted. A hybrid ToA and AoA based centralized MDS method is presented in \cite{Jing2015} for WSNs localization. In \cite{Franco2017} a theoretical generalization for centralized MDS based localization is provided in the presence of few anchors.

\subsection{Semi-centralized MDS Based Localization}

The centralized MDS based localization methods generally have high computational complexity and large localization error for irregular networks. Therefore, many researchers were encouraged to develop semi-centralized (or clustered) MDS methods for WSNs-IoT localization \cite{Shang.Y2, ji, Moore, Gwo, Minhan, Shao2012, Macagano, Chen20013, Dou2010, Wang200015, Saeed2016, Jayasumana2016, Stojkoska, Ghods2017}. Semi-centralized MDS methods are more robust and accurate with low complexity. In recent past the authors in \cite{Imtiaz2016,Imtiaz2017} proposed three dimensional semi-centralized MDS based localization methods for IoT networks.

In semi-centralized MDS based methods, initially, all the nodes in the network are divided into clusters. Different clustering algorithms such as  $k$-means clustering, density-based clustering, or fuzzy clustering can be used for a clustering purpose. Once the network is clustered, the next step is to select a cluster head for each cluster by using various cluster head selection methods such as minimum energy consumption, large number of neighbors etc. Semi-centralized MDS based localization methods consist of following steps:
\begin{itemize}
\item Construction of Local Distance Affinity Matrix: In this step, the cluster head of each cluster computes the shortest path distances for its every member in the cluster and defines a local distance affinity matrix. The local distance affinity matrix for cluster $i$ is defined as $\boldsymbol{D}_i =  \{\hat{d}_{ij}^2\}_{i,j=1}^c$, where $c$ is the total number of nodes in cluster $i$. $\boldsymbol{D}_i$ can be written in matrix form as
\begin{equation}
\boldsymbol{D}_i = \begin{bmatrix}\label{eq: PIM}
0 & \cdots & \hat{d}^2_{1c} \\ \vdots & \ddots & \vdots \\ \hat{d}^2_{c1} & \cdots & 0 \end{bmatrix},
\end{equation}
where $i = 1,2,3....c$ and $j = 1,2,3....c$ are the number of nodes in cluster $i$. 
\item Construction of Local Map for Each Cluster: MDS is applied to the local distance affinity matrix $\boldsymbol{D}_i$ to get the relative position estimate of each node in cluster $i$. Like the centralized MDS, the first step is to double center the local distance affinity matrix i.e. $\boldsymbol{C}_i=-1/2(\boldsymbol{H}_i\boldsymbol{D}_i\boldsymbol{H}_i)$. But the size of the double centered matrix is $c \times c$ instead of $n \times n$. The local double-centered matrix $\boldsymbol{C}_i$ is then decomposed by Eigen value decomposition as
\begin{equation}\label{eq: doublecenter3}
\boldsymbol{C}_i= \mathbf{e}\boldsymbol{\lambda}\mathbf{e}^T,
\end{equation}
where $\mathbf{e}$ represents the eigen-vectors and $\boldsymbol{\lambda}$ are the eigenvalues. Finally, the relative two dimensional positions of the nodes in cluster $i$ are determined from the two largest eigenvalues of $\boldsymbol{\lambda} $ and two largest eigen-vectors in $\mathbf{e}$ i.e.,
\begin{equation}\label{eq: rel11}
\boldsymbol{\hat{P}}_i= {\mathbf{e}_2\sqrt{\boldsymbol{\lambda}_2}},
\end{equation}
\item Stitching of the Local Maps: In this step, the cluster heads communicate with each other to merge their local maps. The local maps are stitched together with the help of inter-cluster nodes, where each inter-cluster node belongs to at least two neighboring clusters. Two neighboring clusters should have at least three inter-cluster nodes for stitching  \cite{Umeyama}. The inter-cluster nodes have different relative coordinates in each cluster, therefore the position estimates of nodes in cluster $i$ after stitching are given as
\begin{equation}
\boldsymbol{\hat{P}}_i = \boldsymbol{A}_i\boldsymbol{\tilde{P}}_s + \boldsymbol{\alpha}_s,
\end{equation}
where $\boldsymbol{A}_i$ is the alignment matrix and $\boldsymbol{\alpha}_i$ is the reconstruction error. For a fixed $\boldsymbol{\hat{P}}_i$, $\boldsymbol{A}_i = \boldsymbol{\hat{P}}_i\boldsymbol{\tilde{P}_i^+}$ that minimizes the reconstruction error $\parallel \boldsymbol{\alpha}_i \parallel^2$, where $\boldsymbol{\tilde{P}}^+_i$ is the  Moore-Penrose inverse of $\boldsymbol{\tilde{P}}_i$, therefore
\begin{equation}
\boldsymbol{\alpha}_i = \boldsymbol{\hat{P}}_i (\boldsymbol{I}-\boldsymbol{\tilde{P}}^+_i\boldsymbol{\tilde{P}}_i).
\end{equation}
The total reconstruction error for all the clusters is given as
\begin{equation}
\sum_{i=1}^c{\parallel \boldsymbol{\alpha}_i \parallel^2}= \sum_{i=1}^c{\boldsymbol{\hat{P}}_i \parallel(\boldsymbol{I}-\boldsymbol{\tilde{P}}^+_i\boldsymbol{\tilde{P}}_i)\parallel^2}.
\end{equation}
Let $\boldsymbol{S}_i$ is a  selection matrix which selects the estimated local positions for the nodes in cluster $i$, such that the Hadamard product of $\boldsymbol{\hat{P}}\boldsymbol{R}_i=\boldsymbol{\hat{P}}_i$ and $\mathbf{\Theta}_i = (\boldsymbol{I}-\boldsymbol{\tilde{P}}^+_i\boldsymbol{\tilde{P}}_i)$ then $\sum_i{\parallel \boldsymbol{\alpha}_i \parallel^2}=\sum{\parallel \boldsymbol{\hat{P}}_i \boldsymbol{S}_i\mathbf{\Theta}_i \parallel^2}$. Decomposing $\boldsymbol{R}\mathbf{\Theta}\mathbf{\Theta}^T\boldsymbol{R}^T$ by  Eigen value decomposition yields
\begin{equation}
\text{EVD}(\boldsymbol{R}\mathbf{\Theta}{\mathbf{\Theta}^{T}}\boldsymbol{R}^T)= \boldsymbol{\Lambda\lambda\Lambda}^T,
\end{equation}
where $T$ is the transpose operator. The global relative coordinates are extracted from the two largest eigen-vectors  $\boldsymbol{\Lambda}$ and the corresponding two eigenvalues $\boldsymbol{\lambda}$ for two-dimensional localization i.e.,
\begin{equation}\label{eq: relative_coord}
\boldsymbol{\hat{P}} = \boldsymbol{\Lambda}\sqrt{\boldsymbol{\lambda}}.
\end{equation}
As the relative global position estimates obtained by \eqref{eq: relative_coord} are not absolute, it is required to transform these relative position estimates into absolute positions. Linear transformations such as Procrustes analysis, Helmert transformation, or principal coordinate analysis can be used to reach the global absolute position estimates.
\end{itemize}

\begin{table*}
\centering
\caption{Comparison of MDS based localization methods for WSNs-IoT\label{tablemdscompare}}{
\begin{tabular}{|p {6cm} |p {3cm}|p {2cm}|p {3cm}|p {2cm}|}  
\hline\hline 
Literature &  Computation &  Complexity &   Topology  &  Accuracy \\ [0.5ex] 
\hline
\cite{Shang, viv, Shi2011,Weber2011,Chen2016,Bingjie2010,Wang20015,Jing2015,Chun2015, Franco2017} & Centralized & $O(N^3)$& Uniform &  1.2-2.21 m  \\
\cite{Shang, viv, Shi2011,Weber2011,Chen2016,Bingjie2010,Wang20015,Jing2015,Franco2017} & Centralized & $O(N^3)$& Irregular &  10-14.5 m  \\
\cite{Shang.Y2, ji, Moore, Gwo, Minhan, Shao2012, Macagano, Chen20013, Dou2010, Wang200015, Saeed2016, Jayasumana2016, Stojkoska, Ghods2017,Imtiaz2016,Imtiaz2017} & Semi-centralized & $O(N k^3)$& Uniform &  0.6-1.2 m  \\
\cite{Shang.Y2, ji, Moore, Gwo, Minhan, Shao2012, Macagano, Chen20013, Dou2010, Wang200015, Saeed2016, Jayasumana2016, Stojkoska, Ghods2017,Imtiaz2016,Imtiaz2017} & Semi-centralized & $O(N k^3)$& Irregular &  6.2-8.4 m  \\
\cite{Jiang2016, Stojkoska2017, MORRAL2016, Costa, Vo2008, Chan2009, Sottile2010,Jamali2012,Kumar2015,Kumar2016,Minghu,Latsu, Shi, Marziani, Zhu, nasir2,Nguyen2016, Nguyen2017, Altinoz2017} & Distributed & $O(N L)$& Uniform/Irregular &  4.3-7 m  \\
\hline
\end{tabular}}
\end{table*}
\subsection{Distributed MDS Based Localization}\label{distributedsec}
Distributed localization methods are required for a wide range of applications. Therefore, fully distributed MDS based localization methods have been recently developed \cite{Jiang2016, Stojkoska2017, MORRAL2016}. In distributed MDS based localization methods every node calculates range measurements to its neighbors and updates its location estimate by minimizing a local cost function \cite{ Costa, Vo2008, Chan2009, Sottile2010}. The steps involved in common distributed MDS based localization are similar to centralized and semi-centralized schemes, but after getting the relative position estimates, the central station sends the relative position estimates to each node, and then each node refine its own position estimate by using iterative position estimators. For example in  \cite{Shi2007} and \cite{nasir2} the authors used the steepest descent and Levenberg Marquardt method, respectively, to refine the MDS based relative position estimations. The update rule for position estimation in \cite{nasir2} is defined as
\begin{equation}
\hat{\boldsymbol{P}}_i^{k+1}=\hat{\boldsymbol{P}}_i^k-(\boldsymbol{J}^T_k\boldsymbol{J}_k+\lambda\boldsymbol{I})^{-1}\boldsymbol{J}_k(\hat{d}_{ij}-f({\hat{\boldsymbol{P}}_i^k})),
\end{equation}
where $\boldsymbol{J}_k$ is the Jacobian matrix given by
\begin{equation}
\boldsymbol{J}_k = \begin{bmatrix}
\frac{\hat{x}_i-x_1}{\sqrt{(\hat{x}_i-x_1)^2-(\hat{y}_i-y_1)^2}} & \frac{\hat{y}_i-y_1}{\sqrt{(\hat{x}_i-x_1)^2-(\hat{y}_i-y_1)^2}}\\
\frac{\hat{x}_i-x_2}{\sqrt{(\hat{x}_i-x_2)^2-(\hat{y}_i-y_2)^2}} & \frac{\hat{y}_i-y_2}{\sqrt{(\hat{x}_i-x_2)^2-(\hat{y}_i-y_2)^2}}\\
\vdots & \vdots\\
\frac{\hat{x}_i-x_L}{\sqrt{(\hat{x}_i-x_L)^2-(\hat{y}_i-y_L)^2}} & \frac{\hat{y}_i-y_L}{\sqrt{(\hat{x}_i-x_L)^2-(\hat{y}_i-y_L)^2}}
\end{bmatrix},
\end{equation}
$\lambda$ is the step length, and $f({\hat{\boldsymbol{P}}_i^k})$ is the error function given as
\begin{equation}
f({\hat{\boldsymbol{P}}_i^k}) = \sum_{l=1}^L\left(\hat{d}_{il}-\sqrt{(x_i-x_l)^2+(y_i-y_l)^2}\right)^2.
\end{equation}
$L$ is the total number of anchors and $\hat{d}_{il}$ is the estimated distance between node $i$ and anchor $l$.

In addition to the low complexity and better accuracy, distributed MDS based methods can also support mobility. Since, in centralized MDS techniques, all the ranging information is collected at the central node which is the bottleneck of the network. In case of mobility, the network topology changes which require the minimization of a new global cost function in real time and therefore, the centralized MDS may not be a practical solution. However, in distributed MDS a local cost function is minimized which does not depend on the global topology of the network and therefore can support mobility. A number of distributed MDS based localization have been developed to determine the location of a moving user/node using different tracking filters. For example, in \cite{Jamali2012} the authors used extended Kalman filter and unscented Kalman filter with MDS to track mobile sensors. A low complexity majorization function with MDS is used in \cite{Kumar2015,Kumar2016} to track mobile sensor nodes. Distributed MDS based localization algorithm is proposed in \cite{Minghu} with noisy range measurements, where the authors assume that the distances are corrupted with independent Gaussian random noise. MDS with different refinement schemes to get better localization accuracy for the sensor nodes location in WSNs has also been proposed in literature \cite{Latsu, Shi, Marziani, Zhu, nasir2}. Recently a Euclidean distance matrix completion method is proposed for MDS in \cite{Nguyen2016, Nguyen2017} to find the map of an IoT network.  Different heuristic methods such as particle swarm optimization, simulated annealing, and genetic algorithms are applied with MDS to determine the location of mobile nodes \cite{Altinoz2017}. 

\subsection{Comparison of Various MDS Based Localization Methods}
Depending on the application scenario each MDS based localization method has its own pros and cons. For example, if the nodes are distributed irregularly then semi-centralized methods have better accuracy than the centralized methods. Also, if the nodes are mobile then distributed MDS based methods are preferred over centralized or semi-centralized methods because the distributed methods have lower complexity and faster convergence.  To compare the centralized, semi-centralized, and distributed MDS based methods for localization, two different scenarios are considered. First, 100 nodes are randomly and uniformly distributed in $100~\times~100~m^2$ square area with four anchors at each corner of the area. The transmission range of each node is 20 m and ranging error is 0.01 m. Based on the transmission range, a multi-hop network setup is established. The single-hop distance between any two nodes exist if they are within there communication coverage; otherwise, the multi-hop distances are calculated. The multi-hop distances are computed using the well-known shortest path algorithm (Dijkstra). Once all the pairwise distances are estimated using the Dijkstra algorithm, then classical MDS is used in the centralized technique to estimate the relative location of each node in the network. Although these relative locations of the nodes can visualize the network, the location of the nodes does not have a global coordinate system. Therefore, with the help of anchors and the coordinate transformation techniques such as Procrustes analysis, global coordinates of the nodes are determined. In the case of semi-centralized approach, the network is first clustered into small sub-networks by using any clustering technique. In this paper, we consider Fuzzy C-means clustering to divide the network. Once the network is divided into clusters, classical MDS is applied locally by the cluster head at each cluster to get the relative coordinates of each node. These clusters are then joined together by using patch stitching techniques to get a complete visual configuration of the nodes. In distributed techniques, first the relative coordinates of the nodes are estimate using the same steps in centralized and semi-centralized techniques, and then these locations are sent to each node to refine their positions.

\begin{figure*}
    \centering
    \begin{subfigure}[b]{0.32\textwidth}
\includegraphics[width=1\columnwidth]{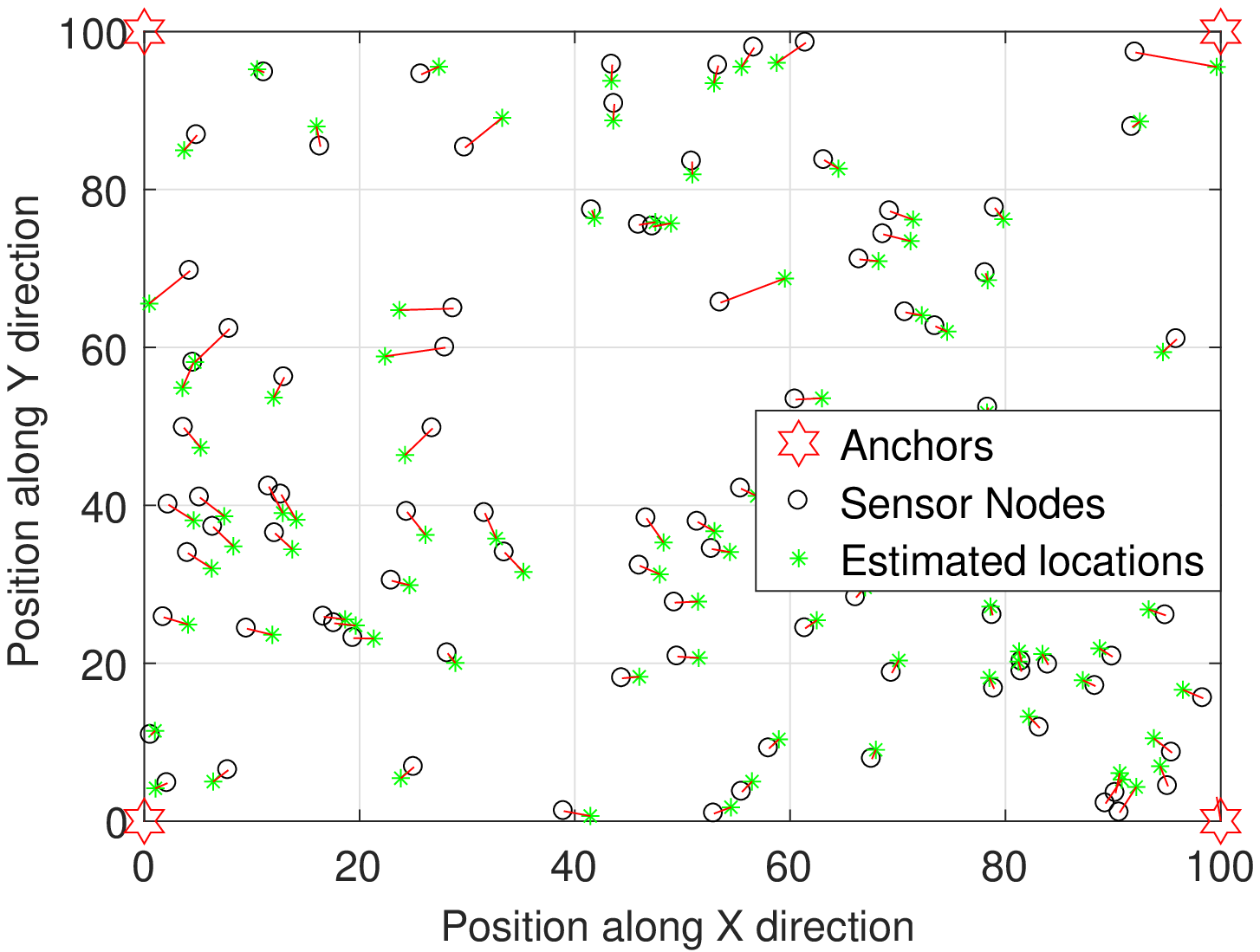}  
\caption{Centralized MDS}
\label{fig:compareunicenter} 
    \end{subfigure}
    \begin{subfigure}[b]{0.32\textwidth}
\includegraphics[width=1\columnwidth]{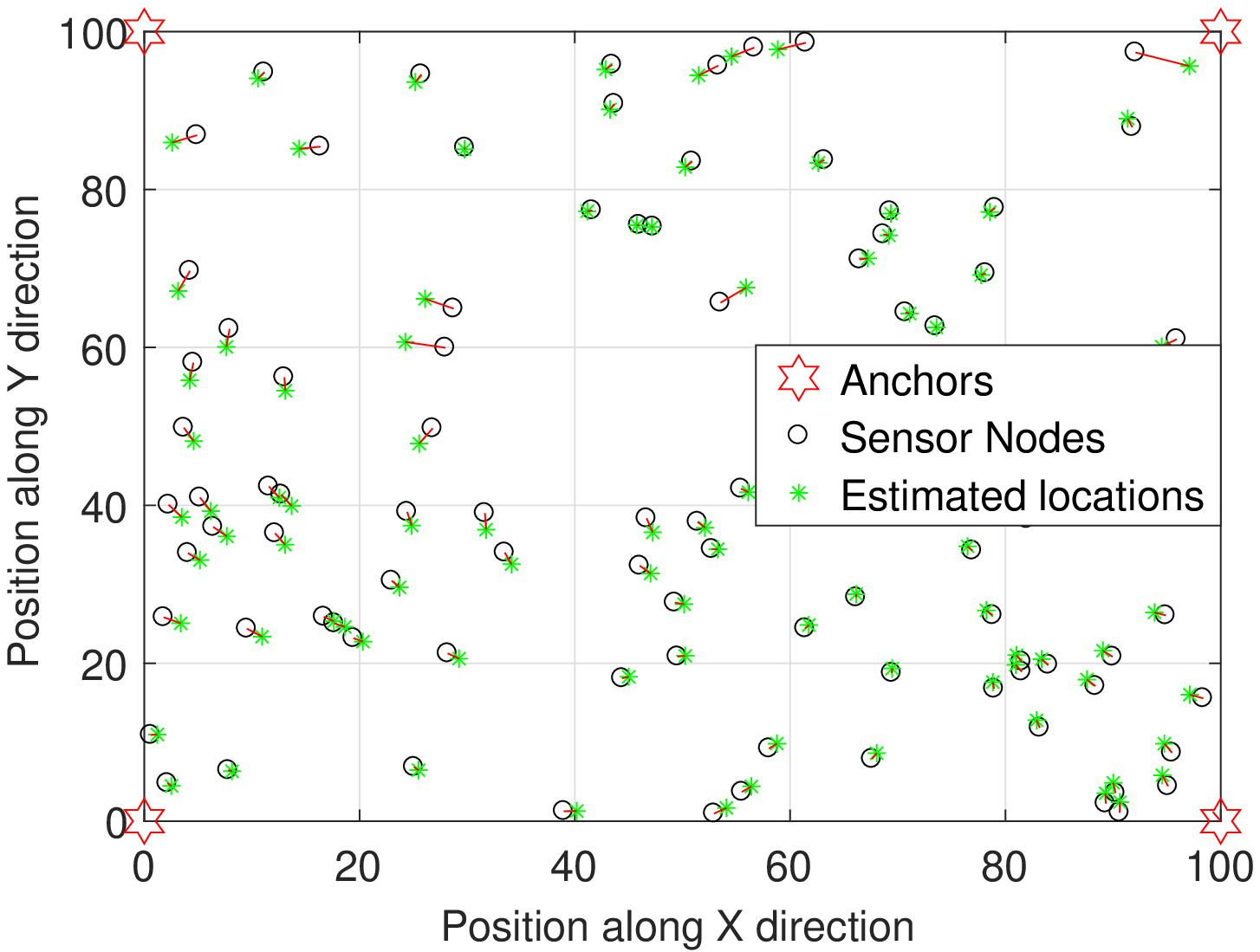}  
\caption{Semi-centralized MDS}
\label{fig:compareunisemicenter} 
    \end{subfigure}
    \begin{subfigure}[b]{0.32\textwidth}
\includegraphics[width=1\columnwidth]{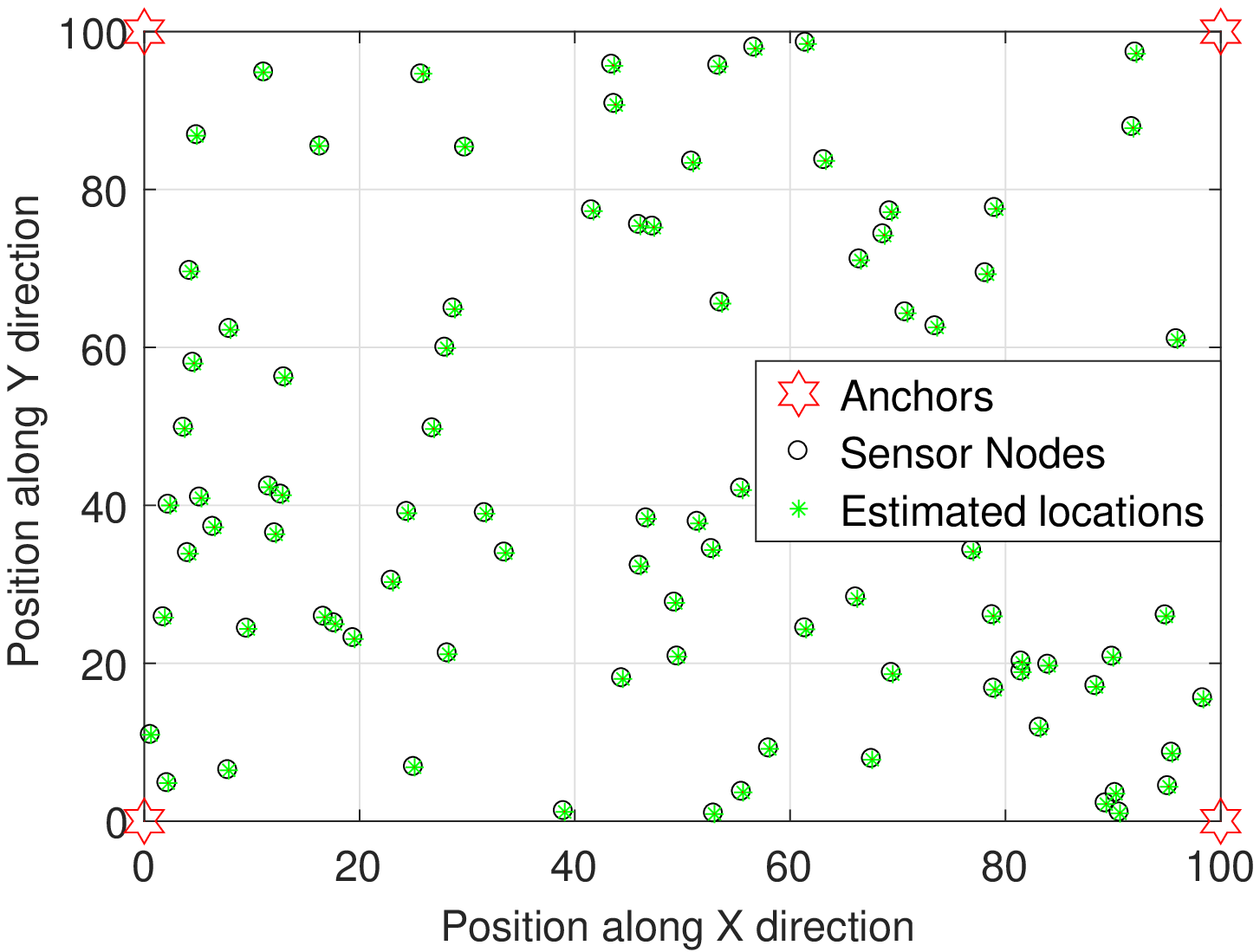}  
\caption{Distributed MDS}
\label{fig:comparedistributed} 
    \end{subfigure}
    \caption{Uniform topology: a) Centralized MDS, b) Semi-centralized MDS, and c) Distributed MDS. }
\label{fig:uniform}
\end{figure*}

Fig.~\ref{fig:uniform} shows that the average localization error of centralized, semi-centralized, and distributed MDS based methods is 2.21 m, 1.2 m, and 0.15 m, respectively. In these figures, the black circle, green asterisk, and red stars represent the actual position of the nodes, estimated position of the nodes, and position of the anchors respectively. The red line shows the localization error for each node.

\begin{figure*}
    \centering
    \begin{subfigure}[b]{0.32\textwidth}
\includegraphics[width=1\columnwidth]{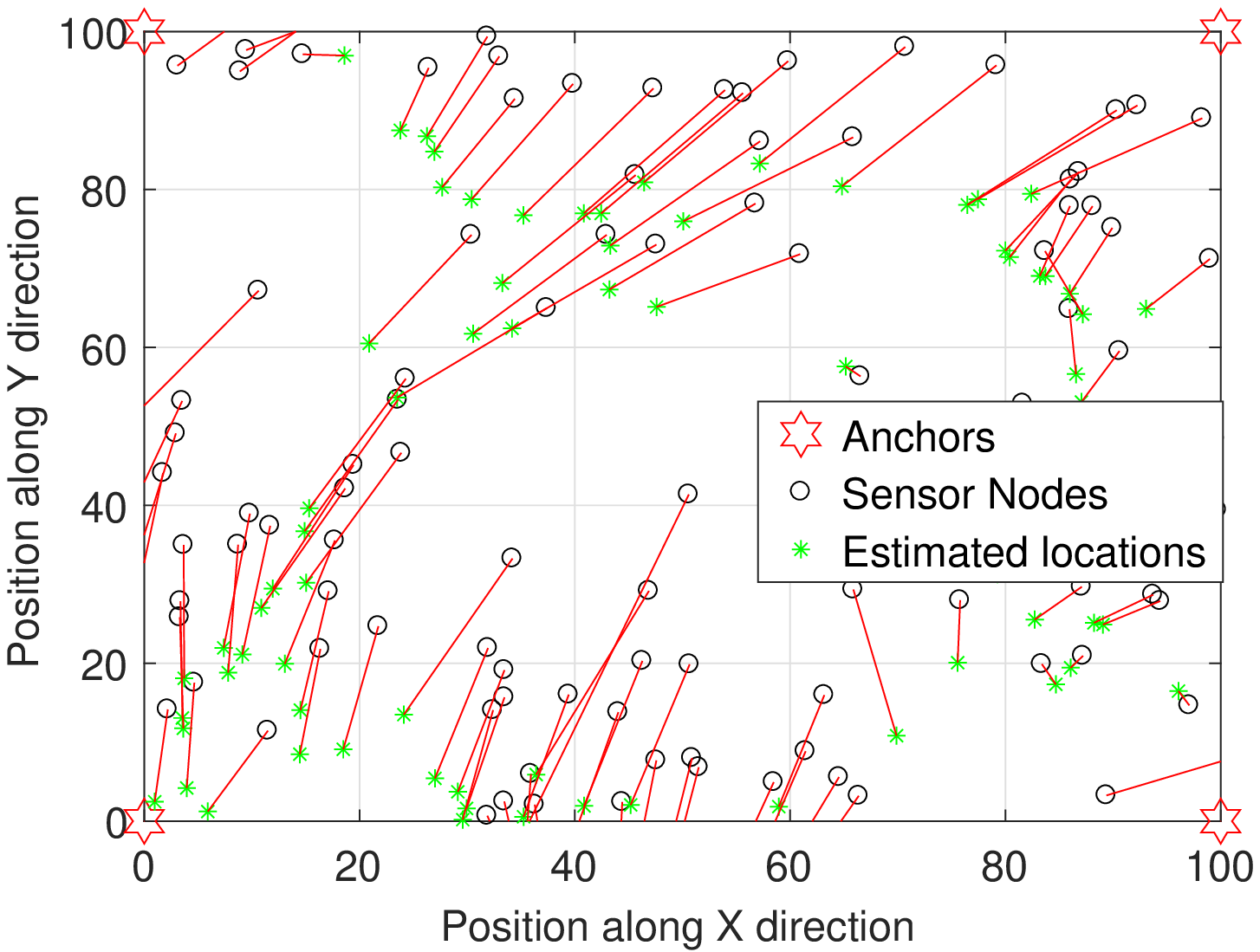}  
\caption{Centralized MDS}
\label{fig:comparenonunicenter} 
    \end{subfigure}
    \begin{subfigure}[b]{0.32\textwidth}
\includegraphics[width=1\columnwidth]{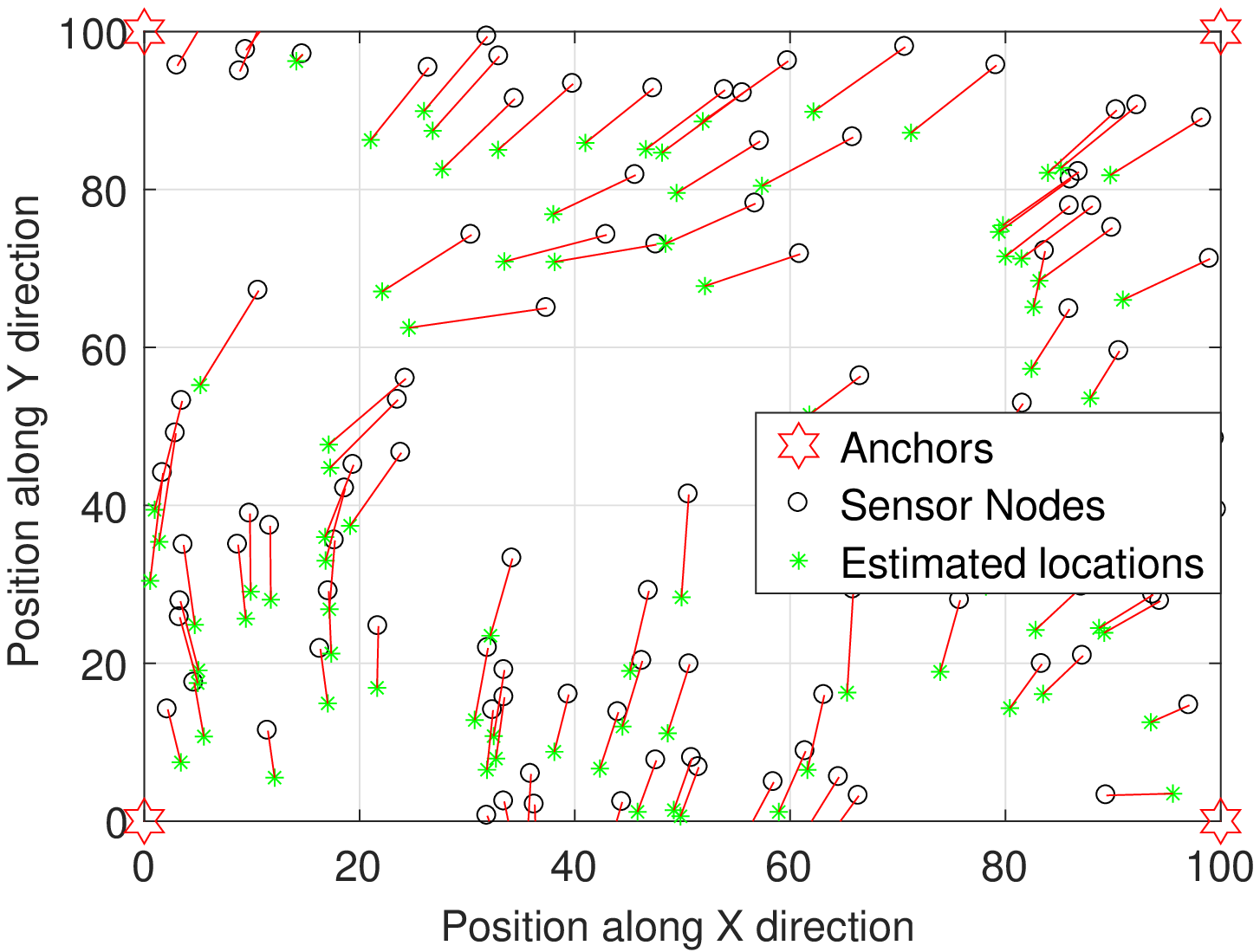}  
\caption{Semi-centralized MDS}
\label{fig:comparenonunisemicenter} 
    \end{subfigure}
    \begin{subfigure}[b]{0.32\textwidth}
\includegraphics[width=1\columnwidth]{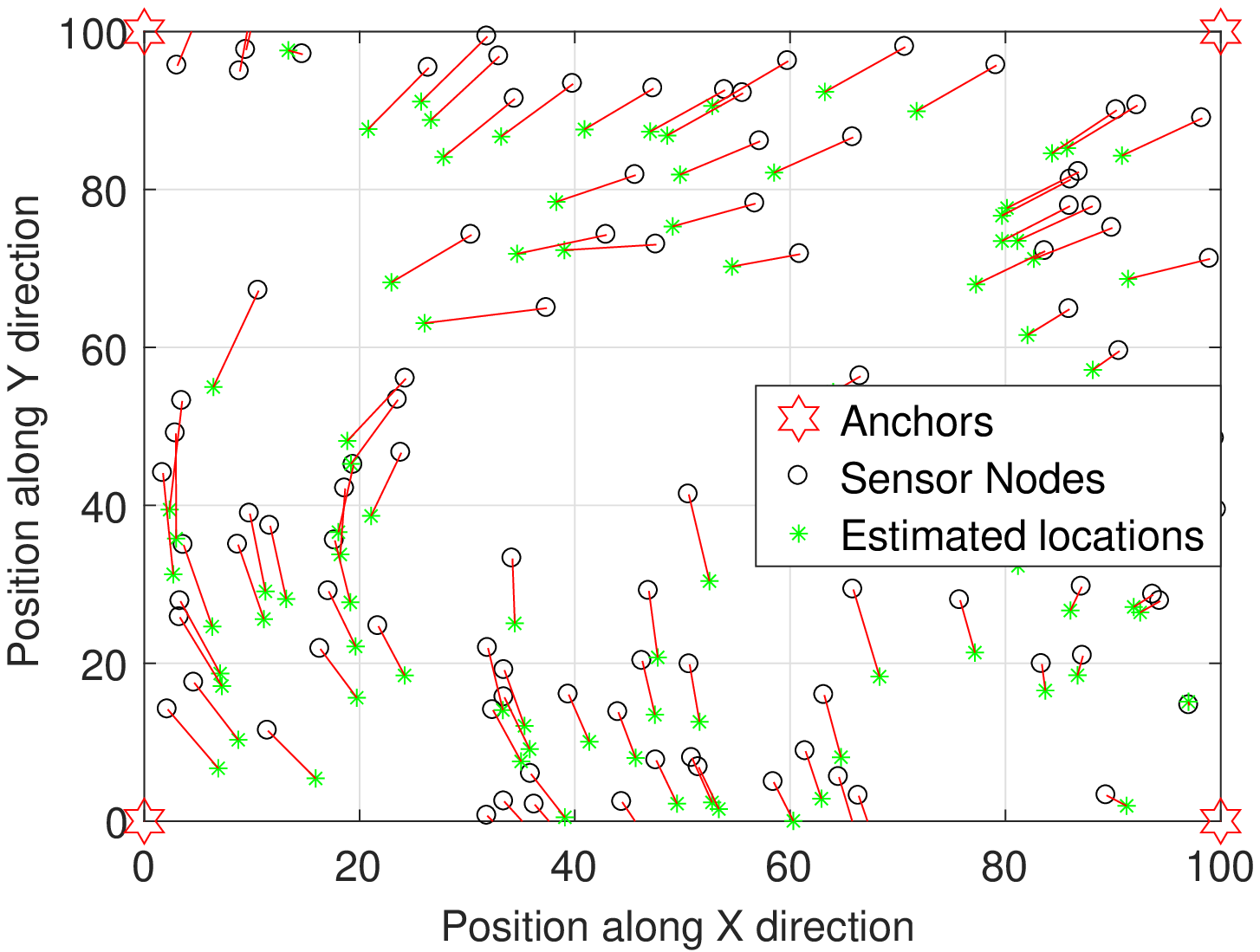}  
\caption{Distributed MDS}
\label{fig:comparenondistributed} 
    \end{subfigure}
    \caption{Non-uniform topology: a) Centralized MDS, b) Semi-centralized MDS, and c) Distributed MDS. }
\label{fig:nonuinform}
\end{figure*}

In the second scenario, 100 nodes are distributed irregularly in $100~\times~100~m^2$ square area. It can be seen in Fig.~\ref{fig:comparenonunicenter} that the centralized MDS based methods have large localization error (i.e., 14.5 m ) in such irregular networks while the accuracy is improved to 8.4 m and 7 m by using semi-centralized and distributed MDS based methods respectively (see Fig.~\ref{fig:comparenonunisemicenter} and Fig.~\ref{fig:comparenondistributed}). The large localization error for centralized and semi-centralized methods are because of the irregularity of the network which causes large shortest path estimation error while in the distributed case the localization accuracy is comparatively better but still its worse than the regular network setup because most of the sensor nodes are not able to get the signals from all of the four anchors. Table \ref{tablemdscompare} summarizes different MDS based localization methods for localization of WSNs, where the network size is $100~\times~100~m^2$. Note that the symbols $N$, $k$, and $L$ in Table \ref{tablemdscompare} represents the number of nodes, number of neighbors, and number of iterations respectively.

\section{MDS Based Localization for Various Networks}\label{sec:MDSlocalization}
This section presents a summary of the literature on MDS based localization for various wireless networks such as WSNs-IoT, cognitive radio networks, and 5G networks.

\subsection{3D MDS Based Localization for WSNs-IoT}
Number of 2D MDS based localization methods for WSNs-IoTs have been discussed in Section \ref{centralizedsec} to Section \ref{distributedsec}. There are various MDS-based accurate localization algorithms proposed for 2D WSNs, but in real-world applications, 3D localization is often needed for better estimation and accuracy. Due to the 3D node deployment and complex environmental factors, the localization algorithms which are effective in 2D environments, have large localization error in the 3D case and therefore cannot be directly applied \cite{Alam}. For example, in 2D WSNs, the location of all the nodes in the network can be computed with the help of three anchors while in 3D networks at least 4 anchors are required. Similarly, the 3D localization technique cannot be directly extended from the 2D solution
by just increasing one extra dimension. There are several problems which can be solved using 2D localization but are much more complex when modeled in
3D space. For example, the distance between any two neighbor nodes in a 2D network is considered as Euclidian distance; however, in a 3D environment, the distances are geodesic rather than Euclidian. Similarly, in 2D networks, the nodes have sufficient connectivity for a given density; however, for the same density, the connectivity is low in the 3D environment due to various obstructions. Hence, localization of WSNs-IoTs in 3D space is an interesting and challenging task. In \cite{Biljana1, Chaurasiya, Wenwei, Biljana2, Saeed2016} the authors proposed 3D localization for WSNs based on MDS. In recent past the authors in \cite{Imtiaz2016, Imtiaz2017} proposed 3D semi-centralized MDS based localization methods for IoT networks.

\subsection{MDS Based Localization for Cognitive Radio Networks}
One of the all-time regulated resources for wireless communication is RF spectrum. From smart-phone users to scanners, from digital TV receivers to door-openers, every wireless device requires an RF spectrum. These ever-increasing demand for RF spectrum leads to deploy the new  concept of dynamic spectrum access \cite{HaykinCR, DongIn2008}. One of the promising technologies to overcome the problem of spectrum scarcity is cognitive radio networks (CRNs) \cite{mitola}.

Many spectrum sensing techniques have been extensively studied in the past decade. Spatial spectrum sensing is  one of the spectrum sensing techniques for unlicensed users to not interfere the licensed users in the spatial domain.  In CRNs, the localization of primary users (PUs) and secondary users (SUs) is
beneficial in order to create an efficient CRN. Since PUs
are not cooperative with SUs in nature, localization of all the users, including PUs,
for the whole CRN is a challenging task.  
In CRNs, localization of PUs and SUs can enhance the system optimization
in following aspects \cite{Nam2015}:
\begin{itemize}
\item Measurements of the spectrum occupancy are precisely performed, 
\item Localization will also determine the reliability of links between SUs,
\item It will help in determining the angle of
arrival/departure of the signal toward PUs, which allows to use beam-forming 
to reduce the interference to the PUs, 
\item Localization will optimize the CRNs, 
thus maximizing the frequency reuse in the space domain, 
\item An optimal SUs network can be modeled based on the location information of PUs.
\end{itemize}
In \cite{andrea, r1, r2}, authors deployed different localization algorithms for CRNs, where they assume that the distances between PUs and SUs are available. As in sensor networks, it is possible to estimate the distances between SUs since they can communicate with each other. But the distance between a PU and a SU cannot be estimated in practice due to the fact that PUs and SUs do not communicate with each other in CRNs. In \cite{YU, 6775388} the authors propose the use of directional antennas for locating the PUs in CRNs. Since the PUs and SUs do not interact, the distances between PUs and SUs is a challenging task \cite{Nasir2015}. In \cite{Nasir2015, Nasir2016, Nasir2017},  MDS based localization methods for CRNs are proposed where the distances are estimated using RSS measurements between the SUs while proximity only (binary) information is considered between PUs and SUs. In \cite{Nasir2015} a centralized MDS based method is proposed to determine the location of PUs and SUs in CRNs. Since the localization accuracy of centralized MDS based method suffers when the network topology is irregular, cluster-based semi-centralized MDS is introduced in \cite{Nasir2016} to determine the location of PUs and SUs. The localization accuracy also depends on the geometry of anchors in the network. An analysis shows that the location of anchors has a large impact on the localization accuracy \cite{Saeed2017coml}.
\subsection{MDS Based Localization for future 5G}
The fifth generation (5G) wireless networks are promising to achieve higher data rates, higher bandwidth low transmission latency.  5G is also considered to be a revolutionary milestone in wireless communication, which will enable lots of new applications including connected cars, IoT with billions of sensors and humanoid robots. Currently, the number of devices connected to the internet are 6.4 billion, in \cite{web1} Gartner predicted that in 2020 the approximate number of devices connected to the internet would reach up to 20.8 billion. In order to support these billions of devices, 5G systems require wide bandwidth which is available in higher frequencies of the radio spectrum \cite{Agiwal2016}. A large number of devices will lead to the deployment of the dense networks in which it is possible to get better-ranging measurements in terms of localization.

In the past, some localization techniques are developed for 3G and 4G networks, but developing localization techniques for 5G is still an open issue \cite{ref1}. LBS are always popular among the users and it is expected to become an essential part of the development of 5G technology. In \cite{14, 60} the authors proposed localization techniques for millimeter waves in 5G. In \cite{17, 25, 49} the authors presented localization schemes for massive multiple input multiple output (MIMO) systems. DoA technique is investigated for 5G in \cite{47}. Similarly, in \cite{60, ali2017} the authors used localization based on RSS. An extended Kalman filter fused with the hybrid DoA and ToA is used in \cite{28} for 5G localization.  A similar technique is presented in \cite{63} in case of a non-cooperative transmitter. 5G Network localization with the dense deployment of users is investigated in \cite{ali2017} using a variant of MDS i.e., Isomap. Localization of static IoT networks in 5G is recently presented in \cite{Emanuel2018} by using centralized MDS.

\section{Applications of MDS Based Localization}\label{sec:applications}
Recently usage of the location based services and applications have seen a drastic increase around the globe. Following are some of the MDS based localization applications.
\subsection{Disaster Management}
The seismic data is difficult to analyze and classical mathematical tools impose strong limitations in unveiling the hidden relationships between earthquakes \cite{machado2013}. MDS based localization is one of the approaches which are useful to get information regarding earthquakes. The maps generated by MDS are intuitive to visualize the complex relationships between seismic events \cite{Lopes2, Lopes2014}. A cluster is formed by similar objects which represent spatially distinguished objects. Earthquake analysis is studied in \cite{Lopes2} and \cite{Lopes2014} by using the data of more than two million seismic occurrences in the period of 1904-2012. The relationship of space-time and space-frequency is used to find the similarity among the events.  The fires caused by different natural factors in forests every year consume vast vegetation areas \cite{Lopes20143}. These forest fires increase the carbon dioxide emission, which contributes to soil erosion and disturbs the water cycle, thus it has a direct impact on the economy of a country. The forest fires in Portugal have been investigated in \cite{Lopes20143} from 1980-2012 by using MDS based methods.

\subsection{Security}
The security conditions can be greatly improved around the globe by using localization. The mobility patterns and different interaction of users can be helpful to determine possible threats for security. Similarly, in war zones, a centralized MDS based localization system is helpful for the military to track its assets and troops which can increase the success of an operation \cite{Julier1999}. The strategic advantage of localization is that the soldiers on the ground pay more attention to the operation and do not worry about the paths for moving forward \cite{Jee2017}. In addition to that, by using centralized MDS localization method the central command can get the global view of the region and can design better plans and strategies.

\subsection{Management and Tracking of Assets}
Management of assets can be achieved by tracking the location of assets which will allow the businesses to perform optimized operations and better inventory management. Distributed MDS based localization and tracking methods can be used to determine the location and track the assets. Localization and tracking based assets management have been extensively studied in the literature  \cite{ lareau2005, Cho2010, GANDHI2010, MOTAMEDI2013, Seol2014}. It is believed that all of the asset management and tracking methods will revolutionize with the advent of IoT. 

\subsection{Internet of Things}
Localization can be of great benefit to IoT networks, for instance, automated services such as handling devices in an office based on users' location.  IoT requires the accuracy of localization in centimeters, therefore, the term used for locating an entity in IoT network is called microlocation \cite{ Zafari2016 }. MDS based localization methods can  be of great use in smart systems such as smart bulidings, smart grids, and smart cities. MDS based localization for IoT networks is still an open research area where very few work exists \cite{Imtiaz2016, Imtiaz2017, Emanuel2018}.

\subsection{Underwater Exploration}
Robust and accurate localization techniques for underwater sensor networks (USNs) is a necessary but challenging task due to the harsh aquatic environment. MDS is used in various works for the localization of underwater sensor nodes. For example, an MDS based localization scheme was proposed in \cite{Saeedtmc} where localization of multi-hop UOWSNs was formulated as an unconstrained optimization problem and solved using the conjugate gradient technique. Similarly, MDS-based localization technique was used in \cite{Nasir2018spawc, Saeedtcom} for three-dimensional underwater optical wireless networks which take into account the outliers in ranging and optimize the anchor's location. Moreover, centralized MDS-based technique was recently used in \cite{Saeed2017, Nasir2018twc} which also considers energy harvesting in the underwater environment to improve the connectivity and localization of the network. All of these MDS-based localization techniques developed of USNs are centralized and thus have high complexity. Therefore, semi-centralized and distributed MDS-based techniques need to be developed for the USNs.

\section{Conclusions}\label{sec:conc}
In this paper, we have presented a comprehensive survey on multidimensional scaling (MDS) technique, MDS based localization in modern wireless communication networks and its applications. This survey covers different aspects of localizations such as global and local localization systems, ranging methods used for localization techniques, a brief discussion of MDS, MDS based localization in cutting-edge technologies (WSNs-IoT, cognitive radios, and 5G networks), and applications of MDS based localization. Besides the detailed study on MDS, the details of different MDS based localization methods are provided along with their use in prospective wireless networks. Centralized MDS based methods are suitable for harsh environments where the localization is carried out at the central station, but for low complexity and better accuracy semi-centralized and distributed MDS based methods are preferred. Also, the possible applications of MDS based localization are provided while the subject remains open to develop accurate and practical MDS based localization methods for current and future wireless networks.


\bibliographystyle{../bib/IEEEtran}
\bibliography{../bib/IEEEabrv,../bib/nasir_ref}

\begin{IEEEbiography}[{\includegraphics[width=1in,height=1.25in]{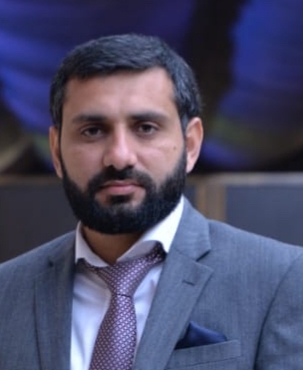}}]{Nasir Saeed}(S'14-M'16-SM'19)
 received his Bachelors of Telecommunication degree from University of Engineering and Technology, Peshawar, Pakistan, in 2009 and received Masters degree in satellite navigation from Polito di Torino, Italy, in 2012.
 He received his Ph.D. degree in electronics and communication engineering from Hanyang University, Seoul, South Korea in 2015. He was an assistant professor at the Department of Electrical Engineering, Gandhara Institute of Science and IT, Peshawar, Pakistan from August 2015 to September 2016. Dr. Saeed worked as an assistant professor at IQRA National University, Peshawar, Pakistan from October 2017 to July 2017. He is currently a postdoctoral research fellow at Communication Theory Lab, King Abdullah University of Science and Technology (KAUST).   His current areas of interest include cognitive radio networks, underwater optical wireless communications, dimensionality reduction, and localization.
\end{IEEEbiography}

\begin{IEEEbiography}[{\includegraphics[width=1in,height=1.25in]{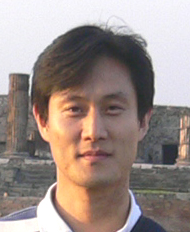}}]{Haewoon Nam} (S'99-M'07-SM'10) received the B.S. degree from Hanyang University, Seoul, South Korea; the M.S. degree from Seoul National University; and the Ph.D. degree in electrical and computer engineering from the University of Texas at Austin, Austin, TX, USA. From 1999 to 2002, he was with Samsung Electronics, Suwon, South Korea, where he was engaged in the design and development of Code Division Multiple Access and Global System for Mobile Communications (GSM)/General Packet Radio Service baseband modem processors. In the Summer of 2003, he was with the IBM Thomas J. Watson Research Center, Yorktown Heights, NY, USA, where he performed extensive radio channel measurements and analysis at 60 GHz. In the fall of 2005, he was with the Wireless Mobile System Group, Freescale Semiconductor, Austin, where he was engaged in the design and test of the Worldwide Interoperability for Microwave Access (WiMAX) medium access control layer. His industry experience also includes work with Samsung Advanced Institute of Technology, Kiheung, South Korea, where he participated in the simulation of multi-input–multi-output systems for the Third-Generation Partnership Project (3GPP) Long-Term Evolution (LTE) standard. In October 2006, he joined the Mobile Devices Technology Office, Motorola, Inc., Austin, where he was involved in algorithm design and development for the 3GPP LTE mobile systems, including modeling of 3GPP LTE modem processor. Later in 2010, he was with Apple Inc., Cupertino, CA, USA, where he worked on research and development of next-generation smart mobile systems. Since March 2011, he has been with the Division of Electrical Engineering, Hanyang University, Ansan, South Korea, where he is currently an Associate Professor. Dr. Nam received the Korean government overseas scholarship for his doctoral studies in the field of electrical engineering.
\end{IEEEbiography}

\begin{IEEEbiography}[{\includegraphics[width=1in,height=1.25in]{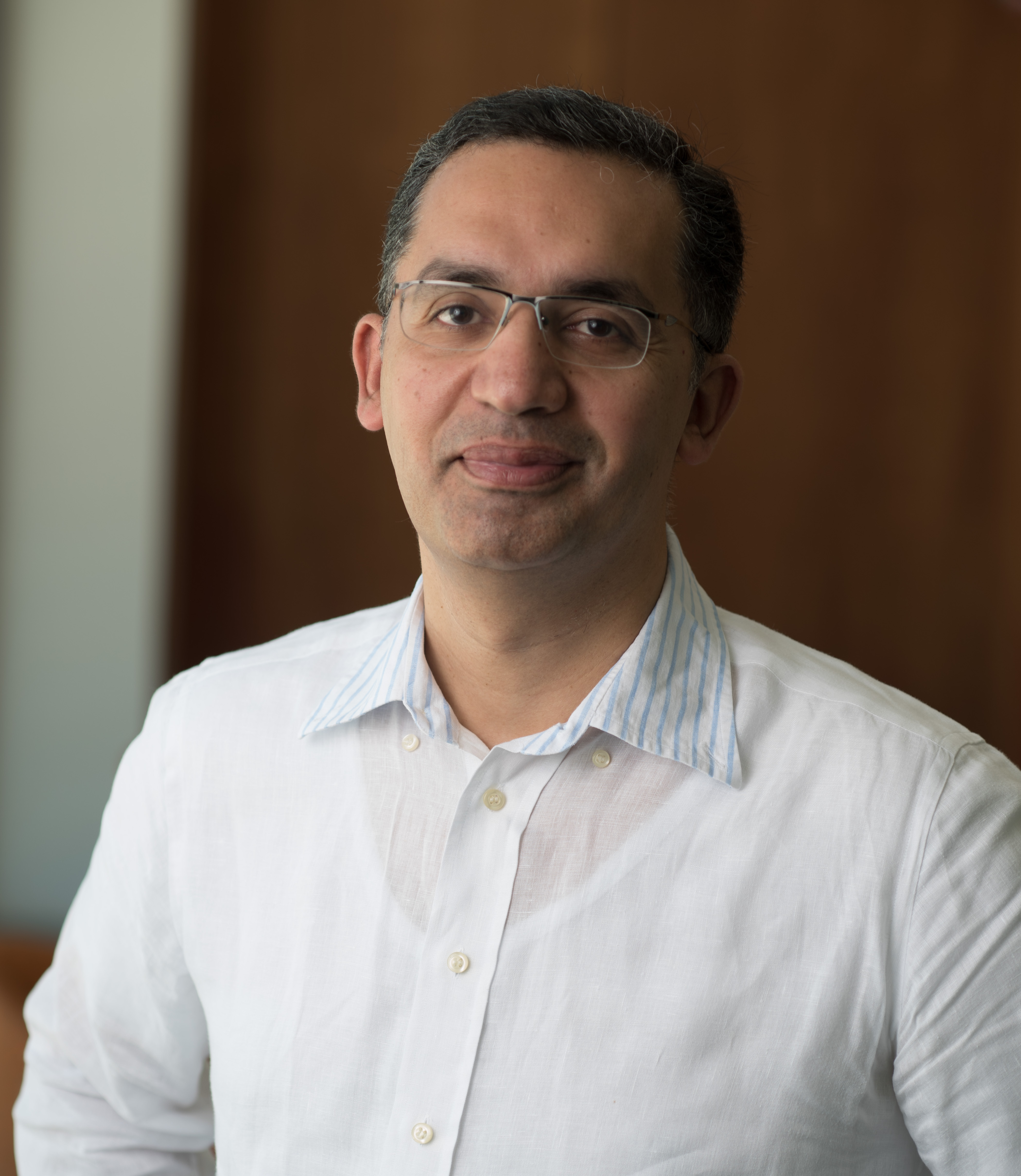}}]
{Tareq Y. Al-Naffouri} (M'10-SM'18) Tareq  Al-Naffouri  received  the  B.S.  degrees  in  mathematics  and  electrical  engineering  (with  first  honors)  from  King  Fahd  University  of  Petroleum  and  Minerals,  Dhahran,  Saudi  Arabia,  the  M.S.  degree  in  electrical  engineering  from  the  Georgia  Institute  of  Technology,  Atlanta,  in  1998,  and  the  Ph.D.  degree  in  electrical  engineering  from  Stanford  University,  Stanford,  CA,  in  2004.  
He  was  a  visiting  scholar  at  California  Institute  of  Technology,  Pasadena,  CA  in  2005  and  summer  2006.  He  was  a  Fulbright scholar  at  the  University  of  Southern  California  in  2008.  He  has  held  internship  positions  at  NEC  Research  Labs,  Tokyo,  Japan,  in  1998,  Adaptive  Systems  Lab,  University  of  California  at  Los  Angeles  in  1999,  National  Semiconductor,  Santa  Clara,  CA,  in  2001  and  2002,  and  Beceem  Communications  Santa  Clara,  CA,  in  2004.  He  is  currently  an  Associate Professor  at  the  Electrical  Engineering  Department,  King  Abdullah  University  of  Science  and  Technology  (KAUST).  His  research  interests  lie  in  the  areas  of  sparse, adaptive,  and  statistical  signal  processing  and  their  applications,  localization,  machine  learning,  and  network  information  theory.    He  has  over  240  publications  in  journal  and  conference  proceedings,  9  standard  contributions,  14  issued  patents,  and  8  pending. 
Dr.  Al-Naffouri  is  the  recipient  of  the  IEEE  Education  Society  Chapter  Achievement  Award  in  2008  and  Al-Marai  Award  for  innovative  research  in  communication  in  2009.  Dr.  Al-Naffouri  has  also  been  serving  as  an  Associate  Editor  of  Transactions  on  Signal  Processing  since  August  2013. 
\end{IEEEbiography}

\begin{IEEEbiography}[{\includegraphics[width=1in,height=1.25in]{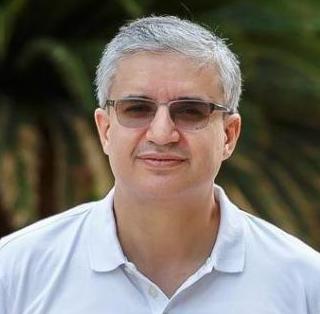}}]{Mohamed-Slim Alouini}
(S'94-M'98-SM'03-F'09)  was born in Tunis, Tunisia. He received the Ph.D. degree in Electrical Engineering
from the California Institute of Technology (Caltech), Pasadena,
CA, USA, in 1998. He served as a faculty member in the University of Minnesota,
Minneapolis, MN, USA, then in the Texas A\&M University at Qatar,
Education City, Doha, Qatar before joining King Abdullah University of
Science and Technology (KAUST), Thuwal, Makkah Province, Saudi
Arabia as a Professor of Electrical Engineering in 2009. His current
research interests include the modeling, design, and
performance analysis of wireless communication systems.
\end{IEEEbiography}

\end{document}